\documentclass[aps,prl,twocolumn,superscriptaddress,groupedaddress]{revtex4}
\usepackage{subfig}
\usepackage{graphicx}  
\usepackage{dcolumn}   
\usepackage{bm}        
\usepackage{amssymb}   
\usepackage{slashed}
\usepackage{graphicx}				
\usepackage{amsmath}
\usepackage{mathtools}
\usepackage{yhmath}
\usepackage{tikz,pgf}
\usepackage{tikz-cd}
\usepackage{comment}
\usepackage{asymptote}
\usetikzlibrary{arrows,backgrounds}
\usetikzlibrary{fit,scopes,calc,matrix,positioning,decorations.pathmorphing}
\usepackage[all]{xy}
\usepackage{yfonts}
	
\usepackage{xcolor}

\newcommand{\Bracket}[1]{\ensuremath{\left\langle#1\right\rangle}}

\DeclareFontFamily{OMS}{oasy}{\skewchar\font48 }
\DeclareFontShape{OMS}{oasy}{m}{n}{%
         <-5.5> oasy5     <5.5-6.5> oasy6
      <6.5-7.5> oasy7     <7.5-8.5> oasy8
      <8.5-9.5> oasy9     <9.5->  oasy10
      }{}
\DeclareFontShape{OMS}{oasy}{b}{n}{%
       <-6> oabsy5
      <6-8> oabsy7
      <8->  oabsy10
      }{}
\DeclareSymbolFont{oasy}{OMS}{oasy}{m}{n}
\SetSymbolFont{oasy}{bold}{OMS}{oasy}{b}{n}

\DeclareMathSymbol{\smallleftarrow}     {\mathrel}{oasy}{"20}
\DeclareMathSymbol{\smallrightarrow}    {\mathrel}{oasy}{"21}
\DeclareMathSymbol{\smallleftrightarrow}{\mathrel}{oasy}{"24}

\begin{document}
\title{On the Renormalisation group, protein folding, and naturalness}
\author{Andrei T. Patrascu}
\address{email: andrei.patrascu.11@alumni.ucl.ac.uk}

\begin{abstract}
I am showing how the ideas behind the renormalisation group can be generalised in order to produce the desired reduction in the degrees of freedom, other that the ones considered up to now. Instead of looking only at the renormalisation group flow, inspiration from optimisation tools for regulators of truncated theories is used to show that there exists another mathematical structure, in the morphisms between various renormalisation groups, characterised by their operations, encoded by means of various regularisation functions. This expands the idea of renormalisation group to a renormalisation category. A group structure exists at the level of those morphisms, leading to new information emerging in the flowing process. Impact on problems like the naturalness and protein folding is being presented briefly. 
\end{abstract}
\maketitle
\section{introduction}
The renormalisation group and the procedure of renormalisation are well known methods in theoretical physics for more than 60 years now [1], [2], [3], [4], [5],  [6], [7]. They are based on a series of assumptions about the physical world that seemed plausible at the time and remain so even now. The main idea behind the renormalisation group is that fewer degrees of freedom can, in certain conditions, be more useful to describe a series of phenomena than all the microscopic degrees of freedom that we know at a certain moment. This computational simplification is fortunate for us, as we can make statistical predictions using far fewer degrees of freedom than what we know to exist "all the way" into the microscopic realm. These assumptions are as follows: 
First, it is assumed that in certain circumstances the dynamics of the system is described preponderantly by something called collective behaviour rather than by the Hamiltonian itself. 
Second, one assumes local interactions (and correlations), third, one assumes the so called "law of corresponding states", where one considers that all fluids have the same equation of state across scales, while the sole distinction would appear from some form of renormalisation of length or energy scales. 
\par All these hypotheses have precisely defined domains of relevance, and beyond those domains we cannot expect a renormalisation group approach to function. The result is that, given these assumptions, different scales can be decoupled from the problem, allowing us to define effective theories that are supposedly only depending on the physics at a given scale. It would be totally nice if things were like that, but, they are not, and probably that's for the better. 

First, one has to make the distinction between collective behaviour and cooperative behaviour. 


What I will call in this article cooperative behaviour is defined by the possibility of correlation and information exchange between scales. Up to now, the basic assumption was that the microscopic degrees of freedom can be replaced with a set of emergent large scale degrees of freedom, while decoupling the behaviour of the small scale degrees of freedom completely from the upper scale. This proved to be too strong an approximation for various situations in high energy physics, like the hierarchy problem, the cosmological constant problem, etc. but it seems completely inappropriate for most of the situations relevant to biology. 
The cooperative behaviour is based on the observation that in large molecules with biological effects and even in the behaviour of biological cells, different scales are not decoupled or separated, but in fact are correlated and communicate with each other. The requirement of a function of a cell at a larger scale, say, in a muscle, gets biochemically translated into signals that determine that cell to have a certain behaviour, to reproduce in a certain way, etc. in order for the larger structure to achieve the desired function. This type of communication and correlation between scales, either through chemical signalling, or simply through the geometry and/or topology of a constituent fitting precisely in a larger structure in order to perform a certain function, is what I call "cooperative" behaviour. Due to the universal applicability of the categorification of the renormalisation group, I believe this "cooperative" behaviour to play a fundamental role in other domains of science, including but not restricted to high energy physics and cosmology. 


In all cases of interest to statistical and quantum field theoretical analyses, what we usually consider is a collective behaviour, not a cooperative behaviour, and we should realise that the two terms are not interchangeable. Because of this confusing identification between cooperation and collective behaviour, a very narrow idea of emergence appeared, one in which collective behaviour of many degrees of freedom amounts to emerging properties that can be described without considering the effects of the other scales from where this behaviour emerges. This principle is almost always contradicted in biology. In biology we deal with large molecules, with various groups that interact, form hydrogen bonds, and twist, bend, or fold in ways that never emerge from a purely collective behaviour. In fact, the very structure of those molecules forbids purely collective behaviour, while not forbidding cooperative behaviour. Moreover, in biology, the cooperative behaviour seems to go across the scales, situations in which amino-acids come together in a certain linear way so that, upon folding, the geometry of the ensemble is such that it fits in a certain other structure are well known [8], [9], [10], [11], [12]. In fact we know of a quaternary hierarchical structure in protein folding [13] which is quite astonishing, given that the cooperation between four scales is required for the function of a specific protein to be realised. We are discussing about large molecules in which many degrees of freedom are bound together in a limited region in which correlation between parts is being noticed due to the special resulting geometry, but in which we do not see the type of collective behaviour we use in other fields like condensed matter, or quantum field theory. In fact, approaches like clustering together atoms in such molecules and treating them as unified structures, eliminating therefore underlying degrees of freedom while creating higher scale, and fewer effective degrees of freedom have been tried, but unsuccessfully so [14]. Clearly the standard approaches of the renormalisation group as we understand it now, does not directly work in such cases [15]. However, recently [16], an AI has been trained on predicting such structures, and it did it quite astonishingly at an accuracy of over 95 percent, being way above any alternatives tried. We may assume that in the learning process, the neural network found some patterns that remained unnoticed to us, but analysing the neural network to find out what patterns those were is rather useless. We all know how little one can gain from solving a problem by just finding interesting patterns while still not understanding the underlying mechanism. 
In any case, while protein folding is not the main aspect of my article, it is definitely a very important example of where things related to the renormalisation group and the decoupling of scales can be improved. 
\par The main question pertinent to the renormalisation group was "how much can we reduce the size of a system without qualitatively changing its properties?" It is generally assumed that this relevant minimal size is related to the correlation length, $\xi$, of the system which clearly depends on the state of the system. When the correlation length is small compared to its underlying components, say, the correlation length is of the order of magnitude of a few atoms only, we can do no better than using quantum chemistry to try to make predictions. I am not going into the depths of the quantum chemistry involved in that situation, but it is worth noting that huge Gaussian bases are involved, and we describe the many electron problem in terms of expansions upon expansions of excited Slater determinants etc. Those approaches are of no consequence anyways, they are extremely time consuming and they are never practical in determining actual results in the domain of large molecules, where people used to rely on other methods that are incomplete in other ways, for example density functional theories [17], etc. 
The domain where the renormalisation group can be applied is where the correlation length is large enough to incorporate a significant number of underlying degrees of freedom. Such systems are generally materials near the critical points and quantum field theories. As, even for a scalar field theory, the field at each point represents a set of degrees of freedom, the continuous nature of the real space gives it a significant (although probably not infinite) number of degrees of freedom per unit of correlation length. Therefore, quantum field theory as used in the standard model is the benchmark application of the renormalisation group, and in fact it was the first place where it was used. Decoupling the scales starting from a standard Lagrangian implementation of a quantum field theory is not quite obvious, and decades of research were required to finally solve the problem. The fact that so many unknown degrees of freedom are simply replaced with very few degrees of freedom which allow finding answers to most of the questions one may ask at a given scale is quite fascinating, however, as we tend to forget when looking at modern quantum field theory, it is important to note that not all questions about quantum field theories have actual answers. We do not for example have any restrictions on the fundamental masses of particles. Even with the flow of the theory's parameters (including mass), we have no constraint at a given length, so that we can fix the initial conditions on the renormalisation group equations, other than empirical evidence. Why particles end up having certain masses and not others is described in various ways, but ultimately, the fundamental answer to the question of what the masses are is not to be found in the standard model and the usual extensions of the standard model do not provide any conclusive answer, although they do hint towards what approaches should be taken. This amounts to a series of other questions emerging in the standard model of particles, like, the hierarchy problem, or the cosmological constant, two probably related problems where the evolution of degrees of freedom across scales becomes important. 
And then comes string theory, with its well known T-duality which reminds us that basically if we look at theories described by strings on largely different scales of compactification, we find the same physics. This is the ultimate coupling between scales, which we still don't know exactly how to interpret in the context of our usual renormalisation group approaches. 
However, it appears that two major problems in science, one in high energy physics, and one in biology, originate in the lack of understanding of the same phenomenon: transition across scales. 
For the sake of completeness in this introduction, let us have a look at what happened in quantum field theory. We started in the early part of the 20th century to try to reconcile special relativity with quantum mechanics, the outcome of this being quantum field theory, basically a theory based on operator-valued distributions on patches of spacetime. The theory worked, also because of the spatially extended nature of the distributions, although we know of methods related to operator product expansions that deal with more extreme situations in which operators are defined on infinitesimally closed regions. In any case, while the formulation was constructed, obtaining results with it was largely impossible, as even the simplest questions beyond the first order in perturbation theory (the Feynman tree diagrams) diverged.
The reason for this was the requirement by any decent quantum interpretation to integrate over all intermediate possible steps when performing an amplitude evaluation. This integration, when inner loops were involved, had to be done over all possible momenta of those loops, leading to integrals defined over all momentum scales. However, very high momentum scales misbehaved, leading us to understand that at some point our quantum field theoretical approach must fail. While we expect to complete that part of the theory with something qualitatively different (possibly strings), it would be strange to assume that we need a perfect understanding of string theory to be able to discuss electron-positron pairs, or the decay of weak particles. In any case, people in the 1930s were not truly able to perform such calculations, and string theory was not available. The solution to this problem appeared in an interesting way: it was realised that there is a scale, $\Lambda$ beyond which we shouldn't care what different degrees of freedom we have, as long as their effective behaviour can be modelled in a way that we can control and understand. We wanted to know where this scale emerges in our calculations, and what its numerical value might be. Given the experience people had with similar reductionist approaches in statistical mechanics and thermodynamics, this approach has been taken and the extreme energy scale going to infinity (or momentum scale) was replaced with a finite but large parameter. It has also been observed where this parameter would appear in the Lagrangian and, fortunately enough, it appeared in a finite number of instances, in the couplings (including masses, charges, and field scales). Those were exactly the divergent structures in the limit in which the cut-off scale was set to infinity again. It was also noticed that finite number of subtractions (or re-scalings, according to the reader preferring subtraction or multiplication based renormalisation) was sufficient to reduce the divergence to finite values. 

Of course, the finite values remained undetermined, but they were in a small number, and that small number of parameters could be fixed by experiment. However, while fixing them at one energy scale, it was soon realised that they weren't fixed in Nature, their values "flowing" when the equations fixing the measurable outcomes of the theory when changing the energy scale of the re-parametrisation ($\mu\rightarrow \mu'$) were imposed. 

Therefore, in order to obtain consistent results for the S-matrix and cross sections, we had to assume modified values for the parameters of the system (say mass, couplings, etc.) as the re-parametrisation scale changed. 
The transformations required to change the energy scale and hence modify the "box" one cuts through our system to take into account only a given sample at a given scale, are implementing different modifications on our coupling constants (and/or parameters) of our theory. 

In a Hamiltonian description, in which each scale is described by a Hamiltonian $H_{k}$, we go from the first term $H_{0}$ towards larger length scales by eliminating clusters of degrees of freedom and replacing them with effective degrees of freedom, we get to the point where for a final Hamiltonian $H_{n}$ the original coupling of the first hamiltonian $H_{0}$, say $u_{0}$ plays no important role anymore. However, this is not the rule. In some cases, some couplings $u_{0}$ are amplified by each scale transformation, say $\tau_{i}$, and hence, following many such transformations, the coupling becomes relevant. This means the degrees of freedom coupled by that coupling also remain relevant at higher scales and have an impact on the overall physics. 

This aspect is often disregarded by assumption, however, it would be important if quantum field theory had a fundamental scalar. It was considered at some point that this is not the case, because the Higgs field would eventually emerge as a composed particle or appear in some multiplet. Another approach was to assume  that supersymmetric partners of the usual particles would contribute to the Feynman diagrams in ways that would eliminate the resulting radiative corrections to the Higgs mass. Neither one of those assumptions turned out to be right. 
Therefore, while the reductionist approach does indeed seem to provide us with some answers, it seems like an incomplete way of looking at nature. We actually do not have a full decoupling of scales and the connections between them, not even in quantum field theory, unless we decide to ignore several important questions like the origin of masses, or the origin and (meta)stability of the Higgs. 
In any case, there is another situation that affects this interpretation. The initial scale Hamiltonian, say, $H_{0}$ is assumed to be local. That means, the dynamics described by this Hamiltonian should permit only couplings directly to the nearest neighbours. If the correlation length is actually larger than the range of interactions of the initial Hamiltonian, then an implementation of the renormalisation group is possible. If however the range of interactions is comparable to the correlation length already for the initial dynamics, then the renormalisation group cannot be directly applied. However, long range initial correlations due to initial interactions do exist in various systems. These effects are due to quantum entanglement. Entanglement is basically an arbitrary length correlation, and therefore in strongly correlated systems the renormalisation group approach is not easily implemented [18]. 
We see therefore several situations in which relations between scales must emerge, the decoupling of the scales as done in high energy physics is not even sufficient for high energy physics questions, and the main assumptions of the renormalisation group approach fail in quite a few cases of fundamental or practical importance. As a rule of thumb, whenever scales seem to interact strongly, in a way that cannot be decoupled or be related, we cannot assume the renormalisation group approach to be complete. Of course, we can have very accurate predictions within the renormalisation group assumptions, but there will always be questions that do not depend on this description. 

\par The main aspect that led me to the idea that a generalisation of the renormalisation group is necessary was the observation of multi-scale cooperation of degrees of freedom so often noticed in biology.

In the same way in which in high energy physics some information seems to be "leaking" from the high energy scales even through our cut-off, by means of relevant operators, relevant degrees of freedom, and/or relevant couplings, in biology we see relevant inter-scale connections almost all the time, as biology is a multi-scale system with clear functions within the scales and across the scales. 
The ideas behind emergence are very practical, but in some instances, it is assumed that there is only a collective behaviour, rather than what we see in biology, where a cooperative behaviour is what generates connections between scales. In physics we assume that the re-scaling doesn't affect the measurable outcomes, for example we derive the renormalisation group equations demanding that the cross sections scale in a proportionate way or remain conserved. In biology we do not have such perfectly defined quantities, but we do talk about functions, in a procedural sense. The functions of a cell are usually oriented towards keeping it alive and functioning, while some of those functions reach out to the next scale, by means of the connections it makes to its environment. However, given the existence of specialised cells, with large molecular structures controlling them, we cannot simply assume that a rescaling of the mechanisms inside the cell lead to some collective behaviour that amounts to its scale-over-reaching functions. There is no clear collective behaviour, but it seems to be a trade-off between the local functions of the cell assuring its survival and the functions the cell is performing within its greater surroundings. At this point biologists lack a tool for identifying how such scale over-arching functions appear, and the usual interpretations of the renormalisation group approach cannot work, as a clear-cut emergence from collective instead of cooperative behaviour cannot be realised in cells. Surely this problem is more stringent in biology than it is in physics. In physics we have done a lot of important discoveries by simply ignoring the questions emerging from the fact that scales are being connected even beyond the cut-off and that there exists a hierarchy problem and the masses of the particles are related to that problem and have no clear explanation. In biology this problem translates in a lack of understanding of the mechanisms that lead to higher scale functions, and in a lack of understanding of protein folding, as one aspect of it. 
Therefore, my view here is that the renormalisation group, although being a very important tool, needs some fundamental generalisations in order to be able to describe Nature properly and more generally. When devising the ideas behind the renormalisation group we oversimplified the problem, being probably fascinated by the predictive power gained by ignoring so many degrees of freedom. The general attitude was to continue on this reductionist path beyond what is fundamentally allowed, while ignoring the questions that appeared and were linked to precisely the situations in which such a reductionist view could not be applied. 

\section{categorical renormalisation}
Writing down effective interactions by expressing effective formulas for couplings is indeed helping us re-organise the dynamical problems of many particle systems into problems in which only effective degrees of freedom need to be used. However, by doing so, we do not only ignore the degrees of freedom that can indeed be taken under an effective control by means of larger scale degrees of freedom, we also ignore higher order interactions and correlations that do persist across scales. 

\par The fact that an artificial learning algorithm detecting patterns in protein folding was able to make predictions about folding of proteins in a much more accurate way, suggests that there is additional structure to the renormalisation group, structure that we ignored and that can be used to simplify other problems, based not on collective emergence, but on more complex, cooperative emergence. Collective emergence is the standard emergence in the case in which all smaller degrees of freedom obey the same dynamics, do basically the same thing, resulting in an emerging higher scale property. Cooperative emergence however is based on a cellular analogy, in which each individual cell has its own function (as described by its own individual instructions in its DNA) that manifests itself as acting internally to support the cell, and externally to look for nutrients, etc. that generate a complex web of networking with surrounding cells. The functions of each cell are not harmonised to behave collectively in only one specified way, but to cooperate to produce very specific effects outside the scale of individual cells. Each cell has a series of independent functions, and each cell is capable of living separately and start acting in favour of its own existence once the connections with its surroundings are severed. So, the question is, how can we integrate this type of behaviour in our mathematical models? 
\par I suggest that in order to do this we need to think of an expansion of the renormalisation group, in which we don't only use transformations that link scales as objects in our (semi)group, but in which we also consider maps between different such groups together with the maps between those maps. In that sense we obtain a 2-category generalisation of the concept of the renormalisation group. A renormalisation group is classified by a series of trajectories of the parameters that flow under the action of the group. Those actions are being represented graphically in the parameter space of the theory (figure 1). Now, such (semi)groups become objects in our expansion of the renormalisation group, where the maps of the category are going to be linking various such objects, hence they will model the modifications of the renormalisation groups according to some other rules that we need to specify. What are those rules? Let's try to find out. In order to model variations within a renormalisation group we have to parametrise the types of renormalisation group equations and their solutions and to allow them to change, by this allowing for a controlled modification of the renormalisation group operations. The dynamics at different scales represented via the respective Hamiltonians are the group elements, that are being connected by the renormalisation group operations. The maps between groups will alter the structure of the transformations and therefore the ways in which the parameters of our theories flow through different scales. With this categorical approach we can design a model in which we can see a dynamics of types of flows and detect the conditions under which those flows change and even become relevant according to outstanding parameters. Basically the procedure of adding maps between elements of a collection of objects is what we call categorification.
\par We do expect that the different types of maps will form a relatively complex structure, but in this case we will be able to study that structure and determine whether it is amenable to some reduction. As we can see, it is not only the degrees of freedom that can be reduced and transformed into effective degrees of freedom, but also the maps between the groups. Therefore we will try to find equations that modify those maps in various ways. With this approach, it is becoming possible to discuss long range correlations and even non-local effects in the theory by just analysing the structure of the maps relating various groups (and their group operations). How would that be possible? Essentially by dealing only with renormalisation groups and a single class of group operations, we restrict ourselves when we start the cutting of "boxes" of our space to regions that eliminate a series of long range correlations and potential non-local interactions (like second order interactions to next or next-to-next neighbours, etc.). However, if we accept maps between those renormalisation group equations, then, within the same categorical structure, we introduce ways in which the renormalisation group operations transform automatically taking into account the longer range correlations not considered in a single-group approach. A renormalisation group category (and its eventual n-categorical expansion) could therefore in principle solve the long range correlation or the strong correlation problem in physics. Moreover, it would be more amenable to calculations involving large scale molecules, and it is probably the structure that the AI is optimising against, when improving the predictions of protein folding. Of course, that this is exactly what the AI is doing is hard to prove and it is only a speculation on my part. If there is also something else involved, I am open to suggestions. What matters here is that this generalisation of trans-scale analysis can be used in figuring out some of the more fundamental aspects of high energy theory and the hierarchy problem. 
I am trying to construct a very general presentation. Due to my desire for simplicity but also because of the intrinsic aspects of the model I will start, unfortunately, with a $\phi^{4}$ theory, but I am sure at some point in the future I will go beyond this approach. The main aspects of the functional renormalisation group follow standard materials as those that can be found in [19]. The main idea of this article is to extend such views by focusing on the renormalisation group from a categorial perspective. 
We start from the partition function 
\begin{equation}
Z[J]=\int \mathcal{D}[\phi]e^{S[\phi]+\int_{r}J\cdot \phi}
\end{equation}
where 
\begin{equation}
S[\phi]=\int_{r}\{\frac{1}{2}(\nabla \phi)^{2}+\frac{r_{0}}{2}\phi^{2}+\frac{u_{0}}{4!}(\phi^{2})^{2}\}
\end{equation}
We have of course $\phi=(\phi_{1},...,\phi_{n})$ the $n$-dimensional real field, $r$ a $d$-dimensional coordinate and $\int_{r}=\int d^{d}r$, $J$ is the external source, coupling linearly to the field and we start by only regarding $S[\phi]$ as $\beta H[\phi]$ where $\beta =1/T$. We can imagine a regularisation cut-off $\Lambda$ in the UV in terms of distance (say a lattice type regularisation). We consider the action above as being "microscopic" i.e. describing physics at lengths comparable to $\Lambda^{-1}$. 
Given a fixed "coupling" $u_{0}$ this model is an $O(N)$ model having a second order phase transition between a disordered phase at $r_{0}>r_{0c}$ and an ordered phase at $r_{0}<r_{0c}$ where $\Bracket{\phi_{r}}\neq 0$ and $O(N)$ symmetry is spontaneously broken. Of course $r_{0}=\tilde{r}_{0}(T-T_{0})$, with the critical temperature given by $r_{0c}=\tilde{r}_{0}(T_{c}-T_{0})$. $T_{0}$ is the mean field transition temperature. 
It is usually interesting to analyse the Helmholtz free energy 
\begin{equation}
F[J]=-T ln Z[J]
\end{equation}
with $J=0$ and the correlation functions such as the two point function 
\begin{equation}
G_{ij}(r-r')=\Bracket{\phi_{i}(r)\phi_{j}(r')}_{c}=\frac{\delta^{2} ln Z[J]}{\delta J_{i}(r)\delta J_{j}(r')}_{J=0}
\end{equation}
Note that $\Bracket{\phi_{i}\phi_{j}}_{c}=\Bracket{\phi_{i}\phi_{j}}-\Bracket{\phi_{i}}\Bracket{\phi_{j}}$.
In this formulation we have a cut-off but we do not understand the connection between the couplings. The standard approach is to define virtual domains of integration around every considered region in the domain of validity of our theory, and step by step, to integrate out the short distance (high energy) degrees of freedom. We create therefore a set of virtual models at each virtual domain, indexed by a momentum $k$ and demand that we can smoothly integrate out the fluctuations of our model as $k$ is lowered from an initial scale $k_{in}\geq \Lambda$ (a scale close to our cut-off) all the way to zero. As the fluctuation of the modes is inversely related to their mass, theoretically what could be done would be to attach a higher, order $k^{2}$ mass to the lower energy modes, suppressing therefore their fluctuations. In this way we "fix" the low energy excitations and consider them "stable" while, at the same time, with respect to them, we dilute the "stability" of higher energy modes. 
This can be achieved by adding a quadratic term to the action of the following form 
\begin{equation}
\Delta S_{k}[\phi]=\frac{1}{2}\int_{p}\sum_{i}^{N}\phi_{i}(-p)R_{k}(p)\phi_{i}(p)
\end{equation}
where the regulator $R_{k}(p)$ is of order $k^{2}$ for $|p|<<k$. The Helmholtz free energy now depends on $k$ and hence $F_{k}[J]=-T ln(Z_{k}[J])$. We construct an average effective action that is scale dependent 
\begin{equation}
\Gamma_{k}[\Phi]=-ln(Z_{k}[J])+\int_{r}J\cdot \Phi - \Delta S_{k}[\Phi]
\end{equation}
where $\Phi=\Bracket{\phi}=\Phi_{k}[J]$ is our order parameter field with $J=J_{k}[\Phi]$. 
We have to make some critical observations about this step. We note that we treated all high energy modes equally, and all low energy modes were "artificially" made significant or, better, persistent. We made the lower energy modes relevant by simply claiming that all high energy modes become equally irrelevant. The problem, of course, is that by doing this we ignore any additional structure that may appear in the high energy domain that may have some impact on the lower energy modes, even though not in the way we expect. After all, the ways in which we expect those modes to have an impact have been integrated out. As I used to ask the first time when I encountered this method, what about the unexpected ways? Maybe in theories like $\phi^{4}$ there are no real unexpected ways in which structure that we integrate out may have an impact, but what about protein folding or the hierarchy problem? There, almost certainly unexpected effects do occur. One way of trying to probe those effects is by analysing the actual limit we are taking here, and see whether we can re-arrange the high energy modes on a surface over which each term goes to zero on a different path, basically allowing for various possible limits. This is indeed possible, but even this type of analysis doesn't fully recover the lost information. Unexpected effects are bound to occur. Otherwise, on the side of the low energy effective theory everything seems working smoothly. We can derive thermodynamic properties from an effective potential 
\begin{equation}
U_{k}(\rho)=\frac{1}{V}\Gamma_{k}[\Phi]
\end{equation}
which is proportional to $\Gamma_{k}[\Phi]$ evaluated in a uniform field configuration $\Phi(r)=\Phi$. This model is $O(N)$ symmetric so we can express this potential as a function of a $O(N)$ invariant say $\rho=\Phi^{2}/2$. $U_{k}(\rho)$ may have a minimum at $\rho_{0,k}$ making the spontaneous breaking of the $O(N)$ symmetry being characterised by a non-zero expectation value of $\Bracket{\phi}_{J\rightarrow 0^{+}}$. Correlation functions can be related to the one-particle irreducible vertices $\Gamma_{k}^{(n)}[\Phi]$ defined as the n-th order functional derivatives of $\Gamma_{k}[\Phi]$. The propagator $G_{k}[\Phi]=(\Gamma_{k}^{(2)}[\Phi]+R_{k})^{-1}$. 
\par The regulator function $R_{k}$ is chosen such that $\Gamma_{k}$ smoothly connects the microscopic action for $k=k_{in}$ to the effective action for $k=0$. 
It is therefore common to require that it satisfies the following properties: first at $k=k_{in}$, $R_{k_{in}}(p)\rightarrow \infty$. This means that all fluctuations are frozen and $\Gamma_{k_{in}}[\Phi]=S[\phi]$. Second, at $k=0$, we take $R_{k=0}(p)=0$ and therefore $\Delta S_{k=0}=0$. This means at low momenta, we take all fluctuations into account and the effective action $\Gamma_{k=0}[\Phi]=\Gamma[\Phi]$ is the effective action of the original model. 
Third, for intermediate momenta, $0<k<k_{in}$, $R_{k}(p)$ is chosen such that it suppresses the fluctuations with momenta below the scale $k$ but leaves unchanged the fluctuations with momenta larger than $k$. 
\par Now, we see that this regulator is based on several assumptions that we have made initially, that are never re-tested for particular systems to see if they apply. We first assume that all information about the high energy modes is encoded in their mass, and because of that, all masses get suppressed equally, or, if not equally, then in a pre-specified arbitrary fashion. There are various types of regulators that work best for specific theories, however, there is nothing natural about making those choices. They are just due to low energy computational convenience. Now, anticipating a bit, given the very specific flows of the couplings in the lower energy domain of effective field theories, across various finite scales, it is strange to expect that all the high energy modes that have been "smoothed" and "integrated out" will only have information carried by their mass values, which we can suppress. What about the distribution of those modes, their configuration in the parameter space of the high end theory? We know of "coupling flow" in the low energy domain, but we expect a clear cut annihilation of those in the high energy domain? Of course, performing the operation as done above results in the renormalisation group equations, but only because we ask for consistent results in terms of the cross sections the theory provides. Constraining the results to something "objective" on one side doesn't however insure that all information on the other side that we simply ignored doesn't affect our results on the side we didn't ignore. We know, however, that there are degrees of freedom and couplings that escape this approach and tend to become relevant or amplified from the cut-off scale into the low energy scales. However, due to our arbitrary approach in the step above, this is pretty much everything we know about them, most of the additional work being done in this field being non-exploratory, but just numerical experimentation. In high energy physics, the main idea was that the UV domain needs to be completed by something else, say, string theory, and that the way in which string theory would affect the low energies is hard to imagine or understand. However, it is possible to construct a better renormalisation group approach in which more potentially useful information is retained, aside of the simple re-scaling. This information must then be analysed itself for relevance, of course. But the type of information that needs to be added must be of a qualitatively different form. Then, on that side we can also have irrelevant constructions that we may average out or eliminate, but in a more consistent way. 
Yes, it is well known that given the proper choices, we can ultimately eliminate the cut-off and remain with only the degrees of freedom relevant to our construction, but that is insufficient for so many theoretical problems. We do get a consistent effective field theory, but its capability of calculating or answering certain questions is drastically diminished. So, what we may ask is: what information is lost by making our specific choices of a Regularisation function? Usually the contextual information, the relationships between the high energy modes, and their potential long range correlations. 
Now let us see what can be done with the system formulated by the restrictions above. Correlation function information about the high energy modes is clearly lost, so, if such correlation can spread through the cut-off, we have no way of noticing it in further calculations, whether we use the RG flow equations or not. The flow equation is the well known Wetterich's formula
\begin{equation}
\partial_{k}\Gamma_{k}[\Phi]=\frac{1}{2} Tr\{\partial_{k} R_{k}(\Gamma_{k}^{(2)}[\Phi]+R_{k})^{-1}\}
\end{equation}
(the trace includes the trace with respect to the space and the $O(N)$ index of the field). The main goal is to determine $\Gamma[\Phi]=\Gamma_{k=0}[\Phi]$ from $\Gamma_{\Lambda}[\Phi]$. Applying successive functional derivatives to this equation we obtain an infinite hierarchy of equations for 1PI vertices. Let's introduce the RG "time" index as $t=ln(\frac{k}{\Lambda})$ and then write 
\begin{equation}
\partial_{t}\Gamma_{k}[\Phi]=k\cdot \partial_{k}\Gamma[\Phi]
\end{equation}
The flow equations for $\Gamma_{k}^{(n)}$ look like one-loop equations but with the vertices being the exact ones $\Gamma_{k}^{(n)}[\Phi]$. If we replace $\Gamma_{k}^{(2)}[\Phi]$ by $S^{(2)}[\Phi]$ in the flow equation, we obtain a flow equation which we can integrate out to give us a one-loop correction to the mean field result $\Gamma_{MF}[\Phi]=S[\Phi]$. The one-loop structure encodes all the information to be obtained by performing all the momentum integrals required at the respective loop level. Therefore for an $l$-loop diagram, a single $d$-dimensional integration has to be carried out instead of $l$ integrations in the standard perturbation theory. Given that by approximating $\Gamma^{(2)}[\Phi]$ with $S^{(2)}[\Phi]$ in the flow equation, any approximation of the flow equation will be one-loop exact. It always includes the one-loop result when expanded in the coupling constants. This means that all results from a one-loop approximation must be recovered from the non-perturbative flow equation. We have to explain clearly how and what it means to allow variation in the renormalisation group transformations, as we already notice that those transformations will be related to the regularisation function and its potential expressions. First, it is well known that the regularisation function will lead to different trajectories in the spaces of the effective actions. If we keep an exact form of the flow equations, we are assured that the final point $\Gamma_{k=0}$ would be the same for all possible trajectories defined by the different choices of the regularisation function. Given that usually approximations in the flow equations must be implemented, there will be some dependence on the shape of $R_{k}$ which is usually used to study the quality of the approximation. However, these different shapes of $R_{k}$ reflect the evolution of the regulator from $\Gamma_{\Lambda}$ up to $\Gamma_{k=0}^{exact}$ and do not take into account the correlations between the fluctuation modes beyond our cutoff. The main role of $R_{k}$ in the traditional approaches is to "fix" or "stabilise" the low energy modes, at the expense of the high energy modes becoming unstable and prone to easy integration. As we can see, none of the $R_{k}$ functions really cared about how the elimination of those modes is made, as long as the results were practical for the low energy domain and the effective theory. That is exactly where an extension to a categorical formulation of the renormalisation group and of the connection between scales needs to start. I will consider a construction in which all possible maps between various $R_{k}$ are being considered, in which I will focus on the ways in which the trajectories of the high energy modes are being eliminated. That means, if we are to add maps to the renormalisation group structure, such that we can lift it to a category, we need to make it so that the high energy modes suppression is done in different ways and results in different geometric configurations. While all of them will amount to suppression of the high energy modes, not all maps doing that will be equally relevant, and therefore we will obtain a hierarchy over such maps as well. This may lead to new composition laws (who knows, maybe even (semi)group like) that will involve those maps. 
The fact that we have $\partial_{k} R_{k}(q)$ in the trace assures us that only momenta $q$ of order $k$ or less will be part of the flow at a given scale $k$ if the decay of $R_{k}(q)$ is fast enough for $|q|>>k$. As we see, we can suppress these modes perfectly if we want, but we never suppress the geometric or topological organisation of the higher energy modes, which is something that can be explored if we allow for maps between such $R$ functions to be analysed and to be given a role in the problem. We have to insure some basic (semi) group property that should be preserved. For example, we wish that for two group elements, the composition is also part of the group, and that this basic rule is reflected into the maps that are part of the new structure of maps defining our category. The operations done on the maps should reflect the operations of the original group. Also, consistency conditions on the operations performed between maps must be insured. Note that by doing this we do not obtain therefore high energy momenta in the low energy calculations. However, we may obtain additional correlations that were not detectable before, that originate in the fact that the higher energy modes may have alternating ways in which they decay to zero in the effective theory. 
If we look back at Wilson's idea of the renormalisation group, we realise that what he wanted was a shell-type integration of fluctuations with a smooth separation between higher and lower energy modes. While such a separation between modes is possible, it is not insured that the maps between the regularisation functions that create this behaviour will interpolate in the same way. It is noted that $R_{k}$ does appear in the propagator $G_{k}(\Gamma_{k}^{(2)}+R_{k})^{-1}$ and cures also the IR divergences that may appear in the vicinity of second order phase transitions. The flow equation doesn't depend on the microscopic action, unlike the original path integral expression for $\Gamma_{k}$. This has prompted many studies aiming at the microscopic dynamics through fixed point analysis and the search for fixed point Lagrangians or Hamiltonians. However, we notice that the additional structure that has been ignored in this approach, is not encoded in the dynamics which remains the same, although maybe unknown at the high energy end, but is encoded in the context, i.e. the geometric distributions of the underlying degrees of freedom, the topology, etc. 
We have to take into account that the benchmark for this problem is the protein folding, where we understand the inner dynamics of molecules, the quantum chemistry is a well developed domain, but resulting configurations, encoded by all possible maps in which one geometry can be mapped into another are extremely hard to predict and to an extent depending exactly on the fluctuations we wanted to consider as small. 
\section{The physicist's mathematics}
This starts badly for mathematicians. What we need is a category representation of something that is essentially a (semi)-group. This is not trivial, although there are various methods by which such an extension can be formally constructed either by adding an identity and allowing for inverses, reducing the semi-group to a domain that behaves like a group, or other methods like internalisation. There will be troubles on the way, no doubt about that, but the idea may have physical implications that out-weight the potential mathematical conundrums which I leave for the mathematicians to solve. 
Therefore, I consider the renormalisation (semi)group to which I conveniently and abstractly associate both an identity and inverses, and start the definitions. 
The operations of a renormalisation group are defined by means of flattening out high energy degrees of freedom by allowing the associated modes to be as un-fixed as possible, while fixing down the low energy degrees of freedom. A renormalisation group will contain all the operations that, followed one-by-one, will transport us from scale to scale. 
\begin{figure}
  \includegraphics[width=\linewidth]{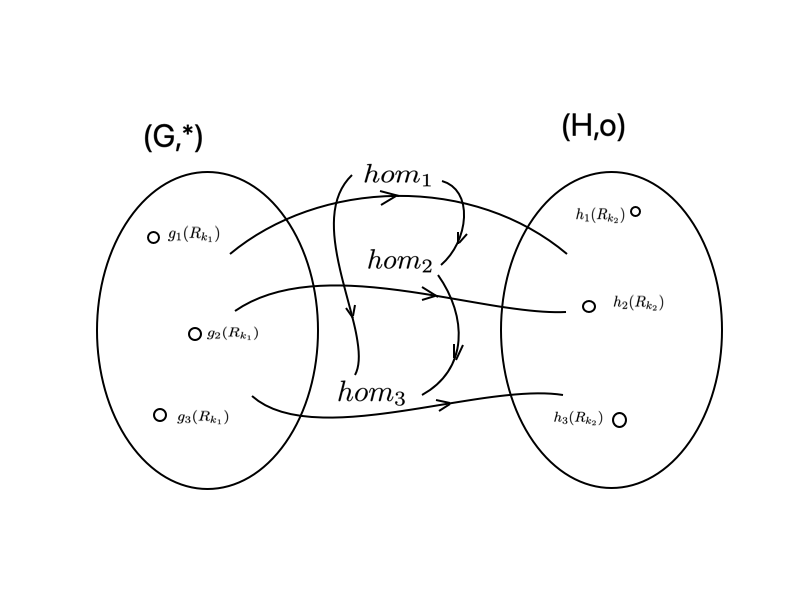}
  \caption{Additional group structure is given to the homomorphism maps linking the renormalisation groups. This amounts to restrictions on the renormalisation group operations and hence on the functions $R_{k_{i}}$ that can reach beyond the UV cut-off and impact the low energy effects. It is as if an additional renormalisation group structure is implemented, but instead of focusing on the groups, one focuses on the homomorphisms between the possible groups}
  \label{fig:gauge1}
\end{figure}

I define two different renormalisation groups as being two renormalisation groups that provide the same exact effective theory starting with regularisation functions $R_{k}$ that behave differently in the high energy domain. Therefore if we have two renormalisation groups $(G,*)$ and $(H,\cdot)$ then $(G, *)$ is characterised by $R^{1}_{k}$ while $(H, \cdot)$ is characterised by $R^{2}_{k}$.
The elements of those groups are respectively $g$ and $h$ corresponding to elements in the groups characterised by the two regularisation functions $R^{1}_{k}$ and $R^{2}_{k}$.
It is worth mentioning that in the classical approach to renormalisation groups, the effective action corresponds to a situation in which all quantum fluctuations have been integrated out, and hence does not depend on choices of the regulator functions. Therefore the freedom of choosing regulator functions is usually exploited in the optimisation of results, given a certain truncation of the effective action. In this article however, by endowing the changes in the regulator functions with a group structure and extending the renormalisation group to a renormalisation category, we are capable of detecting relations between scales that have not been observed previously. We can have $R^{1}_{k}$ and $R^{2}_{k}$ also behaving differently in the low energy domain, although they are expected to provide us with ultimately the same effective theory. Let us consider the two groups related by a homomorphism, say $hom_{1}$ such that each element of $(G, *)$ say, $g_{i}$ will be characterised by $R^{1}_{k}$ and each element of $(H, \cdot)$, say $h_{i}$, will be characterised by $R^{2}_{k}$. The homomorphism condition demands the proper treatment of the neutral element and of the inverses, so that the overall structure is preserved (if we made a choice of an identity and of inverses, no matter how formal) and it also imposes the following relation between group elements 
\begin{widetext}
\begin{equation}
hom_{1}(g_{1}(R^{1}_{k})*g_{2}(R^{1}_{k}))=hom_{1}(g_{1}(R^{1}_{k}))\cdot hom_{1}(g_{2}(R^{1}_{k}))
\end{equation}
\end{widetext}
Now, $hom_{1}(g_{1}(R^{1}_{k}))=h_{1}(R^{2}_{k})$ and $hom_{1}(g_{2}(R^{1}_{k}))=h_{2}(R^{2}_{k})$. To have a group category one has an additional requirement. Moving from one group to the next is forming also a composition law, therefore, the types of transformations leading us from $R^{1}_{k}$ to $R^{2}_{k}$ must be such that the composition rule, given a third group characterised by a third regularisation function, is preserved. Therefore, given two homomorphisms $hom_{1}$ and $hom_{2}$, then if $hom_{1}: G\rightarrow H$ and $hom_{2}: H\rightarrow K$ where we have an additional group $(K, \star)$, then we must have a composition rule across the maps such that $hom_{1}\circ hom_{2} : G\rightarrow K$. 
Let us look in a more practical way at $R_{k}$. This function is defined requiring that it takes the form 
\begin{equation}
R_{k}=p^{2}\cdot r(\frac{p^{2}}{k^{2}})
\end{equation}
with the condition that $r(y\rightarrow 0)\sim \frac{1}{y}$ and $r(y>>1)<<1$. We do not impose any restrictions on the types of functions we can choose outside the domain of interest, and indeed we cannot explore that high energy limit easily. The sole assumption in the usual approaches is that the regularisation function must behave such that the high energy modes can be easily averaged out, with very rapid fluctuations that do not observe any pattern. If we are to assume however that we have an additional category structure over this construction, we cannot truly say that the UV regularisation can be quite as arbitrary as we thought. In fact, while the individual regularisation functions still have a broad domain where they can exist, if we are to consider a categorical approach then we must implement some new restrictions. In particular, combining two transformations between different $R_{k}$ must bring us back to a similar transformation. The only constraint we considered was that in the low energy domain, the effective theory we obtain is the same. We could, for the sake of example also demand that the values at $\Lambda_{UV}$ be the same. In any case, this behaviour together with the requirement that the decay of the small scale large energy fluctuations be equally powerful close to $\Lambda_{UV}$ does restrict our high energy regularisation function and forces us to determine only certain types of exact effective theories, with properties that were not visible before. If for example we demand that the de-amplifying effect of the high energy degree of freedom behave similarly close to $\Lambda_{UV}$ and keeps being smooth in the domain of low energy while at the same time, preserving the composition law of the regularisation functions, hence requiring that the same $\Lambda_{UV}$ behaviour is shared by both functions and their composition, then, several high energy modes are prohibited, requiring a certain minimum distance between them, depending on the rate of drop close to $\Lambda_{UV}$, a requirement that was not there if one just used arbitrary regularisation functions, without a categorical structure. Therefore, we see that this type of categorical approach to the renormalisation group prohibits certain type of UV behaviour, restricting the classes of regularisation functions that can physically be applied. This method therefore would enable us to find correlations in the low energy data that have a controllable way of being determined by the high energy data. 
If we now train for example a large scale deep learning neural network for a problem that has a categorical structure for the maps between maps, then correlations will be automatically detected and used for pattern detection. This aspect however misses in all approaches to the renormalisation group until now. This makes us think, what other structures we missed, by making certain artificial regularisation requirements in quantum field theory or the standard model?
\section{playing with numbers}
Here I will show on some simple cases that demanding a categorical structure for our renormalisation group, it is possible to further restrict the behaviour of low energy effective theories and determine hierarchical properties that were not visible before. 
Moreover, it is interesting to see if a renormalisation (semi) group structure emerges on those "maps of maps" as well, case in which we could combine the transformations between two renormalisation groups, constructing an emerging renormalisation group that depends on fewer "functional" parameters $R^{i}_{k}(p)$. This would create a hierarchy on those maps as well, allowing us for example to fix the Higgs mass by properly taking into account the high energy correlations that are relevant. 
These details are hinted upon even by the early attempts to solve the flow equation. If we have a scale dependent partition function as above
\begin{equation}
Z_{k}[J]=\int \mathcal{D}[\phi]e^{S[\phi]-\Delta S_{k}[\phi]+\int_{r}J\cdot \phi}
\end{equation}
we find the generating functional of the correlated connected functions $W_{k}[J]=ln Z_{k}[J]$ which satisfy the renormalisation group equation
\begin{widetext}
\begin{equation}
\partial_{t} W_{k}[J]=-\frac{1}{2}\int_{r,r'}\partial_{t} R_{k}(r-r')(\frac{\delta^{2}W_{k}[J]}{\delta J_{i}(r) \delta J_{i}(r')}+\frac{\delta W_{k}[J]}{\delta J_{i}(r)}\frac{\delta W_{k}[J]}{\delta J_{i}(r')})
\end{equation}
\end{widetext}
where the derivative is taken at fixed $J$. The scale dependent effective action is
\begin{equation}
\Gamma_{k}[\Phi]=-W_{k}[J]+\int_{r}J\cdot \Phi - \Delta S_{k}[\Phi]
\end{equation}
We obtain the flow equation of this scale dependent effective action at fixed $\Phi$ we consider $J=J_{k}[\Phi]$ with $\Phi_{i,k}[r,J]=\frac{\delta W_{k}[J]}{\delta J_{i}(r)}$. 
Then using 
\begin{equation}
\begin{array}{c}
\partial_{t} W_{k}[J]_{\Phi}=\partial_{t}W_{k}[J]_{J}+\int_{r}\frac{\delta W_{k}[J]}{\delta J_{i}(r)}\partial_{t}J_{i}(r)_{\Phi}=\\
\\
=\partial_{t} W_{k}[J]_{J}+\int_{r}\Phi_{i}(r)\partial_{t} J_{i}(r)_{\Phi}\\
\end{array}
\end{equation}
leading to the Wetterich's equation 
\begin{equation}
\begin{array}{c}
\partial_{t}\Gamma_{k}[\Phi]=\frac{1}{2}\int_{r, r'}\partial_{t}R_{k}(r-r')\frac{\delta^{2}W_{k}[J]}{\delta J_{i}(r)\delta J_{i}(r')}=\\
\\
=\frac{1}{2} Tr[\partial_{t}R_{k}(\Gamma^{(2)}[\Phi]+R_{k})^{-1}]\\
\end{array}
\end{equation}
using that the propagator 
\begin{equation}
G_{k,i,j}[r,r',J]=\frac{\delta^{2} W_{k}[J]}{\delta J_{i}(r)\delta J_{j}(r')}
\end{equation}
is the inverse of $\Gamma_{k}^{(2)}[\Phi]+R_{k}$. Now if we replace $\Gamma_{k}^{(2)}[\Phi]$ by $S^{(2)}[\Phi]$ we obtain the flow equation that provides the one loop result.
It is important to notice that the group structure is what allows us to perform calculations that reach beyond perturbative expansions. It is the way in which the composition rule of the group in the exact and approximate case operate that allows us to go beyond a simple perturbative analysis. If we were to follow the renormalisation transformations in a group, we would reach different scales. The composition of two such transformations, given the group structure, is also a renormalisation transformation and in particular, brings us at the same scale where the two operations would bring us separately. This wouldn't make any difference in an exact description, but in a perturbative description, the group law forces us to perform the scale transition in a very specific way, namely it forces us to correct the errors coming from the perturbative expansion of the evolution in each transformation, by the general over-arching composite transformation. Given that our two perturbative expansions link two different scales of energy (or momenta) in a shell, allows us to systematically use information from the next momentum scale shell to correct our perturbative expansion at the lower momentum shell. In a sense this is amazing. It allows us to "glimpse" inside the non-perturbative structure appearing at the higher energy than what our perturbative approach would allow. Amazing as it is, this has been known for over 60 years now, but it was never assumed that the idea of a renormalisation group may come with more structure, namely in the form of the homomorphism maps between various renormalisation groups. The different groups are parametrised by the transformations we use to move between scales, or, in a sense, by the way we choose to decouple scales. As presented in Figure 1, these transformations also obey a group law. This discussion is obviously imprecise, because in order to apply a group category here, we shouldn't rely on the pure semi-groups we see in the renormalisation group approach. There are however methods that can be used to circumvent this problem, well known to mathematicians, and those details, while definitely modifying the argument I present here, I do not think will affect its overall validity, although some more mathematical work will be required. Even more than what figure 1 presents, the group law associated to the homomorphism structure implies a deeper connection, through which we can gain insight in the non-perturbative or beyond UV aspects of the parametrisations of the group laws, namely of $R_{k_{i}}$. Those functions have been usually chosen to be "convenient" for various computational tools invented. The categorical structure of the group-category however restricts them to have certain properties that allow us a deeper insight into the non-perturbative regime. 
There are different approaches to solving the flow equations as presented here. Two main approaches have become dominant in this field. First, the derivative expansion methods are usually based on postulating certain types of behaviours for $\Gamma_{k}$ considering a finite number of derivatives of the field. The other approach is based on vertex expansions, which try to perform a truncation of an infinite tower of equations satisfied by $\Gamma_{k}^{(n)}$. 
The derivative expansion relies on the regular behaviour of the sale dependent effective action at small momenta $|p|\leq max(k, \xi^{-1})$. A very important aspect of the flow equation is the existence of the term $\partial_{k}R_{k}$ on the right hand side. This amounts to the idea that the integral over the internal loop momenta $q$ is dominated by $|q|\leq k$. However, by looking at this term just in this way, we realise that we do not have any information about correlations existing in the other region that would not appear, not because we equally suppress the lower modes, but because we suppress them in an uncorrelated way. The idea of this article is not to say that the usual renormalisation group approach doesn't work. Instead, it works just fine, and the decoupling of the degrees of freedom works as expected, however, we only deal with the aspects of the high energy theory that we expected to happen. It seems like the additional structure brought by the "maps between maps" has been ignored and that may add a new type of structure that we didn't notice before. It would basically amount to patterns emerging from the way in which the high energy modes are being suppressed via the usual smoothening or averaging out. In any case, if only the group structure is used, we can consider an expansion in the internal and external momenta of the vertices, corresponding to a derivative expansion of the effective action. Now, if we actually expand in terms of derivatives, we obtain the so called Local Potential Approximation (or LPA) in which, at the first level of approximation we can consider that the effective action is fully determined by the effective potential written in terms of the local density, with the field being approximated as constant. 
\begin{equation}
\Gamma_{k}^{LPA}[\Phi]=\int_{r}\{\frac{1}{2}(\nabla\Phi)^{2}+U_{k}(\rho)\}
\end{equation}
It is noted that the derivative term is maintained in the non-renormalised form. Even this first approximation contains vertices to all orders, 
\begin{equation}
\Gamma_{k}^{LPA(n)}\sim \partial_{\Phi}^{n}U_{k}
\end{equation}
Then we can write an equation satisfied by the effective potential that will characterise the above effective action
\begin{equation}
\partial_{k} U_{k}(\rho)=\frac{1}{2}\int_{q}\partial_{k}R_{k}(q)[G_{k,L}(q,\rho)+(N-1)G_{k,T}(q,\rho)]
\end{equation}
where $G_{k,\alpha}(q,\rho) =[\Gamma^{(2)}_{k,\alpha}(q,\rho)+R_{k}(q)]^{-1}$ and $\alpha=(L,T)$ denoting the longitudinal and transversal parts of the propagator $G_{k}(q, \Phi)$ as compared to the order parameter $\Phi$, evaluated for a uniform field. (N-1) corresponds to the number of transverse modes when $\rho$ is nonzero. We can impose the initial condition on the potential as 
\begin{equation}
U_{\Lambda}(\rho)=r_{0}\rho+(u_{0}/6)\rho^{2}
\end{equation}
Using this one would obtain 
\begin{equation}
\begin{array}{c}
G_{K,L}(q,\rho)(q,\rho)=[q^{2}+U'_{k}(\rho)+2\rho U_{k}''(\rho)+R_{k}(q)]^{-1}\\
\\
G_{k,T}(q,\rho)(q^{2}+U'_{k}(\rho)+R_{k}(q))^{-1}\\
\end{array}
\end{equation}
Even in this approximation, we have to note that the regularisation function that characterises the renormalisation group transformation has a bottleneck and is constructed such that it treats higher energy modes equally suppressive. We do not know what happens in a UV completed theory, or, in general we may know, but it is still hard to come up with a way in which those effects will impact the low energy physics. We can however treat the mappings between different renormalisation group operations as a group by themselves, and expand the respective formulas in power series. Those formulas will have a well defined behaviour in the domain of lower energies, but will behave in a rather different way around the cut-off. However, given that we have to preserve the group composition rule, we will have to impose the fact that the composition of two such transformations results in yet another transformation of the same group, so we must obtain yet another homomorphism. 
For one, let's express $R_{k}$ in terms of a Fourier expansion at each level $k$. In that case, the group operation will act on the modes as a composition of Fourier modes. 
Let us therefore discuss of $R^{1}_{k}$ and $R^{2}_{k}$ as expressible in terms of Fourier modes and try to obtain their composition 
\begin{equation}
\begin{array}{c}
(\widehat{R^{1}_{k}\circ R^{2}_{k}})(\xi)=\int R^{1}_{k}(R^{2}_{k}(x)) exp(i\cdot x\cdot \xi)dx=\\
\\
=\int R^{1}_{k}(y)exp(i\cdot (R^{2}_{k})^{-1}(y)\cdot \xi)|det((R^{2}_{k})'(y))|^{-1}dy\\
\end{array}
\end{equation}
Or, otherwise stated if we want to express the composition of the two operations in terms of the Fourier transform of the composition, we obtain 
\begin{equation}
\widehat{R^{1}_{k}(R^{2}_{k})}(\xi)=\int_{\sigma\in G}\widehat{R^{1}_{k}}(\sigma)\cdot P(\xi,\sigma)d\sigma
\end{equation}
where the hat is the Fourier transform. 
In the multidimensional case, we have 
\begin{equation}
(\widehat{R^{1}_{k}\circ R^{2}_{k}})(\xi)=|det((R^{2}_{k})^{-1})|\widehat{R^{1}_{k}}(R^{2}_{k})(\xi)
\end{equation}
with the condition that the inverse of the second transformation should have a unitary determinant for the group rule to be preserved identically. If that transformation between homomorphisms is therefore special i.e. with unit determinant, this implies that all the horizontal transformations at a scale that are permissible should be special.
This is of course a bad example. The reasons why it is bad are relatively obvious: the renormalisation group operation works between scales by "forgetting" small scale degrees of freedom in the process of averaging and taking only the large collective modes into account. This operation is non-invertible and hence discussing in terms of the determinant of the inverse is not a good idea. Assuming there could be a unity and an inverse that could be added could be formally plausible, but physically hard. The particular result here is also most likely not the most generally valid result one could imagine. However, in the same way in which we can associate a category to the Group structure, we can also associate a semi-category to the semi-group structure, where the unity is "forgotten" and the inverse is not always defined. In fact, as will be shown in what follows, it is possible to formally add a unity and the resulting inverse to such a semi-category, and while indeed the two situations are physically distinct, a weaker notion of inverse in the simplest (traditional) representation of the renormalisation group will emerge if one considers a semi-category and the associated homomorphisms obeying the usual group laws. This makes sense given the types of systems that we hope to describe by such methods, namely systems in which specific collaborations between scales as much as within scales do continue to exist. This "collaboration" is not due to the group structure itself but is embedded in the $Hom(F,G)$ structure emerging as a group, encoding the homomorphic maps between $F$ and $G$.
Let us understand the principle behind renormalisation and the renormalisation group intuitively. Let's imagine a theory expressed in terms of a perturbative expansion of some observable, and some parameters, together with the energy scale at which the observable is analysed. We notice that the series expansion is divergent and also that the problem doesn't seem to be related to the series itself, as, in order to arrive at the series we did everything correctly. We correctly observe that our problem appears due to the fact that we made some assumptions about our theory that probably are not valid over the whole spectrum of energies we considered. We also try to determine some of the parameters of the theory and find out the results are generally divergent. We end up making the observation that the problem doesn't lie in the series but instead in the way we parametrised it. As we know, a series can be expanded in various ways, some ways better than others, while in some extreme cases we do not obtain any results at all. The expansion in this case was in terms of a parametrisation that was so bad, the results ended up completely meaningless. The classical approach is then to try to reconstruct the parameters in terms of something else. For example, instead of relying on some fictitious parameters that can only be used internally to our perturbative calculations, we could try to expand the parameters of the theory around a different value, one that we could imagine to be real, given a certain energy scale where it is measured. We therefore expand the parameters in this series around some realistic parameter, and obtain corrections in the expression of the first (unrenormalized or bare) parameters in terms of the realistic parameters. We then go with these parameters in the series expansion of the observable of our theory, and of course, obtain again a series of divergences. The regularisation prescription identifies the divergences by replacing for example some upper limit with a finite value, say the upper cut-off. Other methods, like the dimensional regularisation, can also be used. After this operation is performed, we can re-express the series in terms of the new, physical parameters. What we obtain should be a physical observable at a finite scale (the scale where we perform the measurement), and a physical set of parameters, at the same scale, all depending in some way on the cut-off that we will ultimately send to infinity. But because we constructed now a perturbative expansion for a physical observable and a measured outcome for the parameters at a given scale, those quantities are not divergent. What are divergent are the expressions depending on the cut-off in the limit in which we send this cut-off to infinity, but, from our construction we notice that every time we come with the divergent expansion of the bare parameters in terms of the physical parameters, and insert them in the observable, the $N$th order of the observable expansion divergence is cancelled by the $N-1$ previous terms in the expansion of the bare parameter in terms of the physical parameter. This procedure links different expansion terms and therefore different scales. The common lore of renormalisation group says that all interactions are renormalizable in this way because only a finite number of parameters need to be redefined. This makes the construction reliable and hence we can construct finite quantum field theories. The only exception is gravitation where the number of parameters that need such infinite redefinitions is in fact infinite, and hence an infinite number of measurements would be required to solve this, making the theory non-predictive. 
However, if we go deeper into the problem, we realise that the parameters are not actually infinite, but they are of a different nature. If we go to string theory, we add a lot of information, geometrical and topological, that couldn't possibly be captured by the way we constructed the renormalisation group or the renormalisation prescription. In fact, we know that the flowing of coupling constants is determined by the string theory moduli and their vacuum expectation values, which are just parametrisations of the compactification geometries and topologies. Clearly, the problem is more complex than what we thought, and the reason for this complexity doesn't rely on the inability of the renormalisation group to take into account additional structure, but on the fact that we stopped trying to take into account such structure simply because we never directly observed it in the standard model. Not directly observing it however doesn't mean it's not there. In fact it is, but it appears in terms of the questions nobody is answering now: the hierarchy problem, cosmological constant, Higgs stability, origin of masses, etc. 
The renormalisation group is very suitable for a series of procedures, primarily those of condensed matter systems. There, the parametrisations are simple and usually the emergent phenomena, when scales are being decoupled, are collective. In string theory, what we have as parameters and their flows in the low energy physics, are basically given by the dynamical equations of geometric compactification configurations. And of course, string theory has its own (unspecified) parameters, but far fewer than the "infinite" parameters that need to be solved in a renormalised gravity. It is important to understand the source of this confusion. The actual parameters of the real theory were different from what we considered to be parameters in the low energy structure. In a sense, we had an inter-scale cooperation rather than the collective behaviour we assumed to exist. Each scale has its own structure which may or may not "fit" together properly, and this fitting is not visible if one things only in terms of collective behaviour of constituent particles. In a sense we may say that strings do far more things that particles can do, and that is true. Strings can probe structures of their space in different ways, can detect for example curvature or topological features, they have winding numbers, and, most importantly, the way they interact doesn't only depend on the projected flat space degrees of freedom of a linear string, but also on the ways in which they can wind around all sorts of structures they may encounter. These aspects are external to the view of collective behaviour of particles in a medium (gas) and cannot be explained by such. The only analogy we know of is the folding of proteins and the construction of function across scales that is intrinsic to living organisms. There too, the linear structure of a protein is giving only very little information about its function and role across various scales, but that same protein has a plethora of roles both at its scale and at its surrounding scales. In any case, the problem is real: even if the number of parameters is not infinite in the case of protein folding, it is definitely large, and hence any standard renormalisation group approach would remain impractical. That is, unless we find a method to deal with all this complexity in a unified way. This method seems to be related to a categorical approach to the renormalisation group. 
We generally assume that different groups, separated by different composition laws, are just picks we made for computational convenience we fall into the same mistakes of assuming that it is irrelevant how we change the scales, an assumption valid only in the case in which the underlying degrees of freedom have only a collective, but never a cooperative behaviour. Maybe the name "cooperative" is not the best, there are various types of behaviours we can encounter even in simpler systems like spin glass, or frustrated systems, when in fact we deal with competition in a rudimentary sense (as opposed to the far more complex type of behaviour of proteins and their folding). Anyways, lacking a better word, cooperative behaviour will have to do. 
So, now that we know intuitively where the problem lies, it would be very nice if we had a universal mechanism to solve it. The complexity of the problem is staggering. Like in string theory, in the case of protein folding, many geometrical and topological configurations that may look similar play vastly different roles, and become parts of far more complex systems either via various types of encodings or decodings. It would be definitely hard for a standard algorithm to calculate the effects of all of them in detail. In my opinion there must be a mechanism that is more universal and that controls those effects and their measurable inter-scale relations. The renormalisation group approach relies on several simplifications. The cancellations of divergences between different orders of perturbation expansion is basically a signal for the interplay between scales. If the connection between scales is resolved by eliminating the divergences between the scales, of course, a renormalisation group structure emerges. That process however is not so simple unless one works with very simplified models of condensed matter. We have noticed that the renormalisation group flows of the parameters can behave in very interesting ways, including in ways that manifest chaotic behaviour, etc. and we associated those behaviours with interesting phase transitions, behaviours of spin glasses, etc. 
However, this type of analysis within a single group and following a single type of composition rule, unavoidably misses the higher structure that encodes more complicated situations that do not depend strictly on one simplified (basically collectivist) way of reaching low energy physics. That higher structure, involving competition and collaboration between degrees of freedom within a scale and between scales can be taken into account if one analyses also the dynamics of the maps relating various renormalisation groups that differ for example through the choice of regularisation functions and/or the type of their composition rule. 
Now, those maps between groups have themselves a structure, potentially a group structure. This structure allows for one (or several) hierarchisations of those maps, and hence, a similar renormalisation group approach could be implemented on them. But instead of dealing only with collective subjacent behaviour, what we gain is a universal tool to describe highly complex, maybe even entangled, subjacent behaviour, and the resulting higher scale behaviour. I do not expect the equations of such a dynamics to be solved easily or even to be accessible on any but the best computers, but it could be something that an artificial neural network could be trained to resolve, and it is not excluded that some recent neural networks inadvertently detected such structure. 
Therefore, as expressed in Fig. 1 here, instead of looking only at the vertical structure (which I call the structure following the group elements within each group), it would be very interesting to analyse the horizontal structure as well, the one emerging from the maps (homomorphisms) linking the groups. I will use the terminology "vertical" and "horizontal" with the image of a vertical hierarchy between scales in mind. For convenience Figure 1 represents the morphisms between maps as vertical arrows, but the standard terminology is maintained across the article.
The reason why there is a finite number of parameters in the renormalisation prescription that need to be redefined by means of the same number of finite measurements is because the types of divergences we encounter are having a behaviour across the scales that is relatively easily accounted for by a function determined by a finite number of known interpolation points. If we assume that the large scale behaviour originates from a collective behaviour of structureless components, as is the case in QFT, then this is the case indeed. Yes, there may be complications, which is why we cannot assume that the divergence is fundamentally constant across all scales and only one parameter would be sufficient to measure, but we also have fundamentally a finite number of the same type of parameters, that can all, by successive finite measurements, determine the collective structureless behaviour of the smaller components. This doesn't seem to work for string theory and for protein folding, simply because there, the smallest components have additional structure that leads to a behaviour that links scales in a different way, non-collectivist, but instead competitive and/or collaborative. It is fascinating to note this interesting connection between how complex behaviour of structure-full constituents appears in the two structures we do not truly understand: the connection between protein folding and living organisms, and the connection between the UV completion of our theories, as string theory, and the standard model. 
But let us see what else can be found about large scale structure emergence if we take into account the "maps of maps" that form the categorical underpinnings of group theory. 
It is important to note that the renormalisation group and the renormalisation procedure are linked together and do pretty much the same thing: they transition between scales. However, because we start with an ill defined theory in the form of a perturbative expansion, the initial "renormalisation group" step must be an infinite one, that eliminates the divergences from the workable region of parameters, by redefining the parameters of the theory. Historically, once the scale decoupling background of the renormalisation group has been understood, the same principle was taken and applied independently in theories and situations where the renormalisation procedure was either implicit or not necessary at a first sight. It turned however out many times that indeed there was a need for some form of regularisation and subtraction of divergences and in other cases situations appeared that were not describable by means of the renormalisation group understood traditionally. Strongly coupled systems, entangled systems, etc. are some examples of such cases. 
Of course, all this can be said in a more formal language
A semi-category is defined as a set of objects $|\mathcal{G}|$ such that for all objects $A, B\in |\mathcal{G}|$, an object $\mathcal{G}(A,B)$ is also part of the category. For all objects $A, B, C \in |\mathcal{G}|$ we have a composition law
\begin{equation}
c_{A,B,C}: \mathcal{G}(A,B)\otimes\mathcal{G}(B,C)\rightarrow \mathcal{G}(A,C)
\end{equation}
A morphism of semi-categories will be defined by $F:\mathcal{G}\rightarrow \mathcal{H}$ such that for each object $A\in|\mathcal{G}|$ an object $F(A)\in\mathcal{H}$ exists and for all objects $A, B\in |\mathcal{G}|$ a morphism 
\begin{equation}
F_{A,B}:\mathcal{G}(A,B)\rightarrow \mathcal{H}(F(A),F(B))
\end{equation}
exists. 
Now, with this we can see that additional structure, of the type required to understand the hierarchy problem, doesn't come from the renormalisation group itself, but from its categorical interpretation. A hierarchy over the maps in this (semi)category of renormalisation (semi)groups has a direct impact on the resonances we can detect, shifting them according to the composition laws formed between the homomorphisms between renormalisation (semi)groups. Not only that, but the structure we can add on the homomorphism maps will be useful, given some parametrisation, to the determination of the configurations of proteins after folding. In this sense, the two problems, of protein folding and of hierarchies are related. 
It is well known that the renormalisation prescription, particularly any perturbative renormalisations, but not only, do involve some form of suppression of the degrees of freedom which we consider to be irrelevant at each scale in favour of larger scale degrees of freedom that appear to be more relevant. The renormalisation procedure does this first between a cut-off and a finite scale, replacing the parameters at the cut-off that would diverge otherwise, with the parameters at a specified, well defined scale. This allows the calculation of physical observables below that scale (in energy, or above that scale in length). The observable then calculated at a given scale below our parametrisation scale must be the same, even if we decide to actually re-parametrise the theory at a different scale which is again well defined (so, way below the cut-off we chose initially). The model by which we suppress the degrees of freedom was never considered to be terribly important. The only requirements made upon such model were for it to preserve Lorentz invariance, gauge invariance, and eventually to deal properly with potential IR divergences as well. If however there is hidden structure at each scale, and a different type of structure relating different scales, we cannot continue the same way. The renormalisation group has a certain "self-similarity" note to it. Basically the parameters are flowing in order to maintain the observables at the scale we are interested in the same. We also postulate that the observables of the theory at a given (fixed) scale, which are basically cross sections, etc. are not to depend on the scale where we perform the re-parametrisation. This requirement leads ultimately to the flow of the couplings (or parameters) when our model is parametrised at a different scale. However, the way we deal with quantum field theories is an over-simplification from several points of view. The self similarity of the theory occurs precisely because we decided to eliminate the degrees of freedom (or the modes) at the lower length scales in an indiscriminate manner. 
The alternative would be to take into account all the microscopic details in all complexity and basically put the idea of a renormalisation group aside. But that would be an over-kill. The renormalisation group does help a lot because indeed, in many circumstances the collective behaviour is what is relevant, while individual fluctuations can be suppressed more or less identically and considered as an average. Finally many properties are indeed emergent from this point of view. To deal with structures and properties that do have a role from one scale to the next, either because they are being amplified by the transition between scales or because they are finding new cooperative ways to play a role at different scales is not trivial. In fact, we call such degrees of freedom "relevant" and a very interesting problem is to spot what such degrees of freedom may be and why exactly do they become relevant, as opposed to others being marginal or irrelevant. As said, the extreme approach is to think only in terms of microscopic structure, and that would lead us to quantum chemistry which, as said before, is too complex and complicated to allow for truly emergent properties to be easily found. 
\par The horizontal view I am proposing here is different in the following way. Instead of demanding only the standard Lorentz and gauge symmetry, the cut-off procedure or the regularisation function may become sensitive to other types of structures. Of course, if we knew what structures those were, the problem would be solved in easier ways. Unfortunately neither in the context of the hierarchy problem nor in the context of protein folding do we know that. Allowing for a categorical interpretation of the renormalisation group would however implement a new rule that affects not only the transition between scales, but the changes of the observables at a given scale due to the presence of horizontal structure at the scale where we do the re-parametrisation. 
Basically the renormalisation group says that if we have a quantity $F$ depending on the energy scale where it is measured, call this scale $x$, then in a perturbative expansion, if we know the dynamics (say via some Hamiltonian) if we parametrise it at a scale $\mu$, or at a different scale $\mu'$, we should obtain the same result for $F(x)$. Otherwise stated, $F(x,g_{r},\mu)=F(x,g_{r'},\mu')$. The flow equation will control the change in the parameters (couplings) $g_{r}\rightarrow g_{r'}$. However, this change is not that innocent. At each step we must implement some method of eliminating the degrees of freedom at one scale, and create an effective description at the other. This can be done infinitesimally, hence the $\beta$-functions and their equations, but it always has to consider the underlying properties at each scale. Generally in QFT we are satisfied to consider those properties to be gauge invariance and Lorentz symmetry. However, there are plenty of other properties at each scale, which must be taken into account when the degrees of freedom are being suppressed. For example, if the mode distribution at a given scale has a topological structure, suppressing them uniformly with a function that would eliminate all the higher energy modes the same way, will cancel the topological structure which might as well still have an effect on the distribution of the modes at the next scale. And topology is just a simple example we can think of. Nature is not that "self similar" and "collective" as our calculations would like it to be. But the question is: is there additional order in the features that can emerge at each scale and can interfere with the behaviour of the modes at the other scale? I don't know for sure, but I can at least hypothesise that there is. Therefore I can extend the structure of the renormalisation group in a "horizontal" way, by adding the maps between groups in a categorical sense. This mechanism will allow the procedure to be sensitive to at least some features of horizontal structure. The fact that the homomorphisms linking groups (and hence linking even the composition laws of the respective groups) form themselves a group (or semigroup) is information that can be used. Would it be possible that there is a hierarchical structure of those horizontal connections? 
The existence of such structures is of course known, and one of the main problem in analysing non-perturbative QCD effects via functional methods is related to a better understanding of the cut-off function. However, in all cases up to now, the implementation of structures via
\begin{equation}
\partial_{k}\Gamma[\Phi]=\frac{1}{2} Tr[\partial_{k}R_{k}(\Gamma_{k}^{(2)}[\Phi]+R_{k})^{-1}]
\end{equation}
has been introduced via additions of terms in $\Gamma_{k}$. The next step from the LPA approximation above would have been 
\begin{equation}
\Gamma_{k}^{LPA'}[\Phi]=\int_{r}\{\frac{Z_{k}}{2}(\nabla \Phi)^{2}+U_{k}(\rho)\}
\end{equation}
where the field renormalisation factor $Z_{k}$ is added. This already modifies the regulator function into 
\begin{equation}
R_{k}(p)=Z_{k}p^{2}r(\frac{p^{2}}{k^{2}})
\end{equation}
to allow for a scaling solution of the flow equations and a fixed point at criticality. This leads to a running anomalous dimension 
\begin{equation}
\eta_{k}=-k\partial_{k} log(Z_{k})
\end{equation}
One can continue to a second order estimation of the derivative expansion 
\begin{equation}
\Gamma_{k}^{DE_{2}}[\Phi]=\int_{r}\{\frac{1}{2}Z_{k}(\rho)(\nabla\Phi)^{2}+\frac{1}{4}Y_{k}(\rho)(\nabla \rho)^{2}+U_{k}(\rho)\}
\end{equation}
We have, in addition to the effective potential, another two derivative terms showing that the transverse and longitudinal fluctuations with respect to the local order parameter have different stiffness. The pre-factor of the regulator makes the fixed point condition become identical to the Ward identity for scale invariance. This makes it obvious that the RG fixed point is equivalent to scale invariance. Inserting this "Ansatz" into 
\begin{equation}
\partial_{k}\Gamma[\Phi]=\frac{1}{2} Tr[\partial_{k}R_{k}(\Gamma_{k}^{(2)}[\Phi]+R_{k})^{-1}]
\end{equation}
we obtain a set of four coupled differential equations for the three respective functions and the anomalous dimension. The regulator function is usually considered as a family of functions depending on one or more parameters $\{\alpha_{2}\}$. The optimal value for those parameters is usually determined from what is known in the literature as the principle of "minimal sensitivity", i.e. requesting that locally critical exponents become independent of the parameters of the regulator $\{\alpha_{i}\}$. However, while a lot has been discussed about the role of those functions, the choices are usually arbitrary and the methods of optimisation satisfy rather ad-hoc conditions. I suspect there is more universal approach to this that can be detected by considering an algebraic structure emerging from the (homo)morphisms between such groups. For example, in the $O(2)$ model the transition is driven by topological defects (vortices). The low temperature phase has a line of fixed points that is only recovered after some fine-tuning of the regulator function $R_{k}$. Derivative expansions generally allow all linear symmetries to be implemented in the effective action, and the symmetries of the action are inherited by the physical observables, given that the regulator also satisfies them. To obtain such regulators is usually easy in $O(N)$ but becomes less trivial in gauge theories. We see however, that in all cases, the regulator function should not be dismissed. However, the necessity of fine tuning in the regulator suggests the existence of some inherent structure that has not been taken into account up to now. 
Let us start with a Schwinger functional which has been modified in the following way 
\begin{equation}
e^{W[J,R]}=e^{-\Delta S[\frac{\delta}{\delta J}, R]}e^{W[J]}
\end{equation}
where
\begin{equation}
\Delta S[\frac{\delta}{\delta J},R]=\sum_{n}R^{a_{1},...,a_{n}}\frac{\delta}{\delta J^{a_{1}}}... \frac{\delta}{\delta J^{a_{1}}}.
\end{equation}
is the regulator term, with the property that it should be positive on $e^{W}$ and $\Delta S[\frac{\delta}{\delta J},0]=0$. This amounts, after the usual restrictions to a modification of the kinetic term of the form 
\begin{equation}
S[\phi]\rightarrow S[\phi]+R^{ab}\phi_{a}\phi_{b}
\end{equation}
This amounts to a modification of the propagation of the field. It is interesting to note that such an operator adds source terms for composite operators in the Schwinger functional. As we have seen, we introduced by means of $e^{-\Delta S}$ a source term for $\phi_{a}\phi_{b}$ with current $R^{ab}$. Obviously the regulators can have bosonic and fermionic components which demands particular attention when commuting with $\frac{\delta}{\delta J}$. Mixed bosonic-fermionic terms are also possible. 
The usual general normalised expectation values 
\begin{equation}
I[J]=\Bracket{I[J, \phi]}
\end{equation}
can then be obtained as 
\begin{equation}
I[J]=e^{-W[J]}I[J,\frac{\delta}{\delta J}]e^{W[J]}
\end{equation}
can be re-written with the regularisation terms as
\begin{equation}
I[J, R]=e^{-W[J,R]}e^{-\Delta S[\frac{\delta}{\delta J}, R]}I[J, \frac{\delta}{\delta J}]e^{W[J]}
\end{equation}
We can of course transfer the regulator complete to $I$ leaving us with 
\begin{equation}
I[J, R]=e^{-W[J, R]}I[J, \frac{\delta}{\delta J}, R]e^{W[J,R]}
\end{equation}
A general flow describes how the theory reacts to the variation of the source $R$ and upon integration provides a solution for the theory. If we are given a correlation function $\mathcal{O}[J,R]$ then the flows are being described by derivatives of the correlation function with respect to $R$
\begin{equation}
\delta R^{a_{1},...,a_{n}}\frac{\delta \mathcal{O}[J,R]}{\delta R^{a_{1},...,a_{n}}}
\end{equation}
We can define a parameter space that characterises the different regulators, for example by means of $R(\xi)$. These are trajectories in the space of regulators and therefore describe trajectories linking different theories. If the theories are only defined in terms of truncations it is common to use equations 
\begin{equation}
\delta R^{a_{1},...,a_{n}}_{\perp}\frac{\delta \mathcal{O}[J,R]}{\delta R^{a_{1},...,a_{n}}}_{R_{stab}}=0
\end{equation}
to find stable flows. 

One approach used before was to restrict the regulators to a set $\{R_{\perp}\}$ given the operators providing a regularisation of the theory at some physical cut-off scale $k_{eff}$. If truncations or approximations are being used (as is usually the case) this relation can be used to adjust the flow towards a final effective sector closest to the physical one. In fact, this equation doesn't generally lead to a single stable regulator $R_{stab}$. The search usually must imply variations perpendicular to the direction of the flow (between effective theories at different scales), and hence must be in the space defined by $\{R_{\perp}\}$. Unsurprisingly, this involves not only truncation effects (which are well known to exist and are usually optimised by various techniques) but also real effects due to the actual physical interconnectivity between the scales, that cannot directly and naturally be taken into consideration by the renormalisation group prescription. At this point we may consider additional structure over the operations $R^{a_{1},...,a_{n}}$ and ask what would happen if the relations linking different renormalisation group flows given by different types of regularisations are not only restricted by the usual criteria of providing for example renormalised flows, etc. but also are restricted by the existence of a group structure that could reveal patterns that appear at the transition between scales, and even originating from beyond the cut-off? Let us therefore consider the horizontal transitions between renormalisation groups provided by the changes of the group laws due to the construction of such perpendicular regulators $R$. This implies two effects being part of the variations of $R$. On one side, the truncation effects or the effects resulting from limited series expansions, that lead to imperfect flow convergence to effective theories. Various optimisation constructions can help in this case, assuming the types of different theories are only truncations we wish to re-parametrise and define such that non-perturbative information becomes accessible. Here however, there is more to it. As $\delta R^{ab}$ has the ability to probe the space of possible theories, the idea that the renormalisation (semi)group can be extended in a categorical sense gives more insight into what this transformation can be. Let us give $\delta R^{ab}$ a group structure as identified by some generators $T=T^{\xi}$ such that $R^{ab,\xi}=R^{ab}T^{\xi}$. Moreover, the fact that this horizontal group structure of homomorphisms is linking two renormalisation groups provides us with a similar structure. The same type of scale transition must be found over the maps as well. 
In the case in which the regulator is zero, $R=0$, we have $\mathcal{O}[J]=\mathcal{O}[J,0]$. Total functional derivatives with arbitrary variations of the regulator functions $\delta R^{ab}$ are said to probe the space of possible theories given by $W[J,R]$. We represent the dependence of the regulators in a first stage as depending on a parameter defined to be in $k\in[\Lambda, 0]$ with $R(k=0)=0$. Such one parameter flows are obtained by considering variations of the type 
\begin{equation}
\delta R=dt \partial_{t} R
\end{equation}
where $t=log(k/k_{0})$ provides us with a logarithmic cut-off scale. We can pick $k_{0}=\Lambda$ leading to $t_{in}=0$. Using the flows derived by this $\delta R$ leads us to correlation functions $\mathcal{O}_{k}$ that connect the initial context at $\Lambda$ to the correlation functions $\mathcal{O}=\mathcal{O}_{0}$ of the full theory. The procedure continues by means of a successive integration of momentum modes of the fields $\phi$, turning $k$ into a momentum scale. Therefore we consider the well defined initial condition for large regulators $R\rightarrow \infty$ and we take into account regulators that lead to an IR regularisation with IR scale $k$, $k\in[k_{in},0]$. A most intuitive choice would then be
\begin{equation}
\Delta S[\phi]=R^{ab}\phi_{a}\phi_{b}
\end{equation}
given $R=R(p^{2})\delta(p-p')$. 
Given some well-behaving conditions general one parameter flows can be obtained with such regulators given the condition that $R(k=0)=0$ such that the endpoint of such a flow is the full theory. We obtain 
\begin{equation}
e^{W_{k}[J]}=e^{-\Delta S_{k}[\frac{\delta}{\delta J}]}e^{W[J]}
\end{equation}
where
\begin{equation}
\Delta S_{k}[\frac{\delta}{\delta J}]=\Delta S[\frac{\delta}{\delta J}, R(k)]
\end{equation} 
Our theory acquires a scale and therefore 
\begin{equation}
I_{k}[J]=e^{-W_{k}[J]}\hat{I}_{k}[J,\frac{\delta}{\delta J}]e^{W_{k}[J]}
\end{equation}
where
\begin{equation}
\hat{I}_{k}[J,\frac{\delta}{\delta J}]=e^{-\Delta S_{k}[\frac{\delta}{\delta J}]}\hat{I}[J,\frac{\delta}{\delta J}]e^{\Delta S_{k}[\frac{\delta}{\delta J}]}
\end{equation}
Of course, we also have to represent the constraints and conditions that may be involved in this in terms of the newly introduced scale, for example if a symmetry existed, of the form $I[J]=0$, such relation has to be constructed as $I_{k}[J]=0$ when a cut-off is present. 
The flow equation for cut-off dependent quantities like $I_{k}$, expressed as $\partial_{t}I_{k}$ with $t=log(k)$ and $J$ fixed allows the computation of $I[J]$ in general if the initial context at $\Lambda$ is well defined. Let us denote the flow as 
\begin{equation}
\hat{F}[J,\frac{\delta}{\delta J}]=\partial_{t} \hat{I}[J,\frac{\delta}{\delta J}]
\end{equation}
considering the notation 
\begin{equation}
\Delta \hat{I}=[\partial_{t}, \hat{I}]
\end{equation}
The $t$-derivative acts on everything for fixed $J$ therefore
\begin{equation}
\partial_{t}\hat{I} G[J]=(\partial_{t} \hat{I})G[J] + \hat{I}\partial_{t}G[J]
\end{equation}
After the introduction of a regulator and scale, we will re-express these functions as $F_{k}$, $I_{k}$, $\Delta I_{k}$. The complete Schwinger functional $W[J]=W_{0}[J]$ does not depend on $t$ and hence $\partial_{t}W=0$ which means that, using 
\begin{equation}
I[J,R]=e^{-W[J,R]}e^{-\Delta S[\frac{\delta}{\delta J},R]} \hat{I}[J,\frac{\delta}{\delta J}]e^{W[J]}
\end{equation}
we obtain 
\begin{equation}
F_{k}=\Delta I_{k}
\end{equation}
Using 
\begin{equation}
[\partial_{t}, R^{a_{1},...,a_{n}}\frac{\delta}{\delta J^{a_{1}}}...\frac{\delta}{\delta J^{a_{n}}}]=\dot{R}^{a_{1},...,a_{n}}\frac{\delta}{\delta J^{a_{1}}}...\frac{\delta}{\delta J^{a_{n}}}
\end{equation}
and inserting $\hat{F}$ into 
\begin{equation}
\hat{I}_{k}[J,\frac{\delta}{\delta J}]=e^{-\Delta S_{k}[\frac{\delta}{\delta J}]}\hat{I}[J, \frac{\delta}{\delta J}]e^{\Delta S_{k}[\frac{\delta}{\delta J}]}
\end{equation}
we obtain 
\begin{equation}
\hat{F}_{k}=(\partial_{t}+\Delta S[\frac{\delta}{\delta J},\dot{R}])\hat{I}_{k}
\end{equation}
and inserting this into 
\begin{equation}
I_{k}[J]=e^{-W_{k}[J]}\hat{I}_{k}[J,\frac{\delta}{\delta J}]e^{W_{k}[J]}
\end{equation}
we get
\begin{equation}
e^{-W_{k}}(\partial_{t}+\Delta S[\frac{\delta}{\delta J}, \dot{R}])e^{W_{k}}I_{k}=\Delta I_{k}
\end{equation}
and using $\frac{\delta}{\delta J}e^{W_{k}}=e^{W_{k}}(\frac{\delta}{\delta J}+\frac{\delta W_{k}}{\delta J})$ leading to 
\begin{equation}
(\partial_{t}+\dot{W}_{k}+\Delta S[\frac{\delta}{\delta J}+\Phi, \dot{R}])I_{k}=\Delta I_{k}
\end{equation}
where $\Phi=\Bracket{\hat{\Phi}}_{J}$ is the expectation value of the operator coupled to the current 
\begin{equation}
\Phi_{a}[J]=W_{k,a}[J]
\end{equation}
Putting $I_{k}=1$ and $\Delta I_{k}=0$ we get 
\begin{equation}
\dot{W}_{k}+(\Delta S[\frac{\delta}{\delta J}+\phi, \dot{R}])=0
\end{equation}
and more detailed
\begin{equation}
(\partial_{t} + \sum_{n}\dot{R}^{a_{1},...,a_{n}}\cdot (\frac{\delta}{\delta J}+\phi)_{a_{1}}\cdot ... \cdot(\frac{\delta}{\delta J}+\phi)_{a_{n-1}}\cdot \frac{\delta}{\delta J^{a_{n}}})W_{k}[J]=0
\end{equation}
which is the flow equation of the Schwinger functional connecting the Schwinger functional to a combination of connected Green functions $W_{k,a_{1}...a_{n}}$.
In general, the expansion of the theory is either done perturbatively, or, lacking some small parameters, truncations are being used. Such truncations create artificial regulator dependences in the final theory, and in general there exist various tools by which regulators can be chosen such that the regulator dependences in the final theory are minimised at any given truncation. This is basically also the main reason this formalism has been developed to begin with. However, such a tool allows us to move across regularisation functions, parametrised by one or more parameters, leading to mechanisms that allow us basically to move between renormalisation group operations of different renormalisation groups. Our goal here is not to optimise the regularisation prescription to certain truncations in order to make the final theory as independent of the regularisation as possible maintaining all the benefits. Instead I will consider the possibility that not all dependence on the regularisation function is irrelevant and that in reality additional structure visible in the high energy domain can be detected in the low energy domain of our effective theories and in the respective renormalisation group flows. How can this happen, given that ultimately our theory is indeed independent of the regularisation function chosen? In fact, only the direct dependence in one single flow is eliminated, not the relations that can be constructed between flows that differ by our chosen parameters. While all lead to a final effective theory in absolute terms, that means, if we had access to a full theory that doesn't require a perturbative approach or a truncation, as long as we do not know what that theory should be in absolute terms, we can indeed demand group law structures over the different regulator functions such that, the same absolute theory is recovered (the equivalent of the self-similarity requirement in the standard renormalisation group), but in which variations between various group operations is taken into account. Making such demands results in a re-scaling of the low energy modes in ways that were not detectable previously.
In general the flow equations imply successive integration of degrees of freedom in the general quantum theory. We can imagine that the current $J$ and the regulator $R$ can couple to $\hat{\Phi}(\hat{\phi})$ that are not necessarily fundamental fields $\hat{\Phi}=\hat{\phi}$. The regulator introduces a new scale, modifying the renormalisation group properties of the theory. At any infinitesimal flow step $k\rightarrow k-\Delta k$ there is a natural $k-dependence$ reparametrisation of the degrees of freedom. 
Theories depend on several parameters, including couplings and masses. There exists a response of the theory to an infinitesimal change of the scale of the full theory. There could exist a dependence of the couplings and the currents on this scale $g=g(s)$ and $J^{a}=J^{a}(s)$. We can construct a general linear operator
\begin{equation}
D_{s}=s\partial_{s}+\gamma_{g\;\;\;\j}^{i}g_{i}\partial_{g_{j}}+\gamma_{J\;\;\;\;b}^{a}J^{b}\frac{\delta}{\delta J^{a}}
\end{equation}
with 
\begin{equation}
D_{s}W=0
\end{equation}
with $\gamma_{J}$ the anomalous dimensions. Given $\gamma$ independent of $J$ we only consider linear dependences of the currents. Non-linear relations can be reduced to linear ones via coupling of additional composite operators to the currents. We can construct similar identities for general N-point functions 
\begin{equation}
D_{s}W_{,a_{1}...a_{N}}=(D_{s}W)_{,a_{1}...a_{N}}-\sum_{i=1}^{N}\gamma_{J\;\;\;\;a_{i}}^{b}W_{,a_{1}...a_{i-1}ba_{i+1}...a_{N}}
\end{equation}
using 
\begin{equation}
[D_{s},\frac{\delta}{\delta J^{a}}]=-\gamma_{J\;\;\;\;a}^{b}\frac{\delta}{\delta J^{b}}
\end{equation}
Now, we define an operator $\hat{F}$ again but with the more general differential operator 
\begin{equation}
\begin{array}{c}
\hat{F}=D_{s}\hat{I}\\
\\
\Delta \hat{I}=[D_{s}, \hat{I}]\\
\end{array}
\end{equation}
Now, imposing $D_{s}W_{k}=0$ we obtain $F_{k}=\Delta I_{k}$. For $\hat{I}=1$, $\Delta I_{k}=0$ and using the commutator term of the regulator with the differential operator 
\begin{equation}
[\gamma_{J\;\;\;\;b}^{a}J^{b}, R^{a_{1}...a_{n}}\frac{\delta}{\delta J^{a_{1}}}]=-n\gamma_{J\;\;\;\;b}^{\;\;a_{1}}R^{b a_{2}...a_{n}}\frac{\delta}{\delta J^{a_{1}}}...\frac{\delta}{\delta J^{a_{n}}}
\end{equation}
Then we have
\begin{equation}
[D_{s},\Delta S[\frac{\delta}{\delta J},R]]=\Delta S[\frac{\delta}{\delta J}, (D_{s}-\gamma_{J})R]
\end{equation}
where we use the contraction
\begin{equation}
(\gamma_{J}T)^{a_{1}...a_{n}}=\sum_{i=1}^{n}\gamma_{J\;\;\;b}^{a_{i}}T^{a_{1}...a_{i-1}ba_{i+1}...a_{n}}
\end{equation}
Now, with 
\begin{equation}
\begin{array}{c}
\dot{R}\rightarrow (D_{s}-\gamma_{J})R\\
\\
F_{k}=\Delta I_{k}\\
\\
\end{array}
\end{equation}
we obtain 
\begin{equation}
(D_{s}+\Delta S_{1}[\frac{\delta}{\delta J},(D_{s}-\gamma_{J})R])I_{k}=\Delta I_{k}
\end{equation}
where $\Delta \hat{I}=[D_{s},\hat{I}]$. The operator $\hat{I}$ produces an $s$-scaling encoded in the term $\Delta I_{k}$ and the term $\Delta S_{1} I_{k}$ encodes the additional scaling produced by the operator $\Delta S$. Therefore this equation includes the general scalings in the presence of the regulator. 
For example for an N-point function $I_{k}^{(N)}=\Bracket{\phi_{a_{1}}...\phi_{a_{N}}}$ we can calculate the RG equations introducing $D_{s}=D_{\mu}$ and this implements the RG rescaling in the full theory. The regulator must not destroy the RG invariance of the theory and therefore the commutator
\begin{equation}
[D_{s},\Delta S[\frac{\delta}{\delta J},R]]=\Delta S[\frac{\delta}{\delta J}, (D_{s}-\gamma_{J})R]=0
\end{equation}
vanishes. The RG equation can then easily be obtained for $I_{k}$ as
\begin{equation}
(D_{\mu}+N\gamma_{J\;\;\;a_{1}}^{b})(I_{k}^{(N)})_{ba_{2}...a_{N}}=0
\end{equation}
given that $\Delta I_{k}$ induces a scaling $N\gamma$ of the N-point function. 
In the case of quadratic regulators we obtain
\begin{equation}
(D_{s}+\frac{1}{2}[(D_{s}-\gamma_{J})R]^{ab}\frac{\delta}{\delta J^{a}}\frac{\delta}{\delta J^{b}}+\phi^{a}[(D_{s}-\gamma_{J})R]^{ab}\frac{\delta}{\delta J^{b}})I_{k}=\Delta I_{k}
\end{equation}
where
\begin{equation}
[(D_{s}-\gamma_{J})R]^{ab}=D_{s}R^{ab}-2\gamma_{J\;\;\;c}^{a}R^{cb}
\end{equation}
We assume that $\gamma_{J}$ does not mix fermionic and bosonic currents, and that $R^{ab}=(-1)^{ab}R^{ba}$. For a quadratic regulator, the general s-scaling of the Schwinger functional is derived by using 
$\hat{I}=\frac{\delta}{\delta J^{a}}$ to obtain 
\begin{equation}
\Delta I_{k}=-\gamma_{J\;\;\; a}^{b} W_{k,b}
\end{equation}
and we have also 
\begin{equation}
D_{s}W_{k,a}+\gamma_{J\;\;\;a}^{b}W_{k,b}=(D_{s}W_{k})_{,a}
\end{equation}
Now, if we use 
\begin{equation}
(D_{s}+\Delta S_{1}[\frac{\delta}{\delta J}, (D_{s}-\gamma_{J})R]) I_{k}=\Delta I_{k}
\end{equation}
we obtain 
\begin{equation}
[D_{s} W_{k}+\frac{1}{2}(G_{bc}+\Phi_{b}\Phi_{c})[(D_{s}-\gamma_{J})R]^{bc}]_{,a}=0
\end{equation}
and upon integration
\begin{equation}
D_{s} W_{k}=-\frac{1}{2}(G_{ab}+\Phi_{a}\Phi_{b})[(D_{s}-\gamma_{J})R]^{ab}
\end{equation}
where $G_{ab}=W_{,ac}$ being the full field dependent propagator. This can be used in order to express the flows of 1-particle irreducible quantities or of the effective action as functional of the mean fields $\Phi_{a}=W_{,a}$ and hence to replace the dependence on the current $J$ and its derivative $\frac{\delta}{\delta J}$ in terms of the mean field and its derivative $\frac{\delta}{\delta \Phi}$. We can therefore use it as a transformation matrix for
\begin{equation}
\frac{\delta}{\delta J^{a}}=G_{ab}\frac{\delta}{\delta \Phi_{b}}
\end{equation}
The equation for $W_{k}$ therefore gives us the response of the theory to a general scaling including the flow 
\begin{equation}
(\partial_{t} + \Delta S^{a}[J, \dot{R}]\frac{\delta}{\delta J^{a}}) I_{k}=\Delta I_{k}
\end{equation}
and of the RG rescaling. The equation
\begin{equation}
(D_{s}+\Delta S_{1}[\frac{\delta}{\delta J},(D_{s}-\gamma_{J})R])I_{k}=\Delta I_{k}
\end{equation}
could be re-written in a form depending on the 1-particle irreducible quantities and fields. The action of the operators $D_{s}$ on the functionals $F[\phi]$ is given by 
\begin{equation}
D_{s}=(s\partial_{s}+\gamma_{g\;\;\;\;j}^{\;\;i}g_{i}\partial_{g_{j}}+\gamma_{\phi\;\;\;\;a}^{\;\;b}\Phi_{b}\frac{\delta}{\delta \Phi_{a}})
\end{equation}
Writing $\tilde{I}_{k}=I_{k}[J(\phi)]$ we get $D_{s}\tilde{I}_{k}$ as 
\begin{equation}
D_{s}\tilde{I}_{k}[\phi]=D_{s}I_{k}[J]+((D_{s}-\gamma_{J})J[\phi])^{a}G_{ab}\tilde{I}_{k}^{,b}[\phi]
\end{equation}
These expressions can be lifted to include general variations by making the change
\begin{equation}
D_{s}\rightarrow D_{R}=\delta R^{a_{1}...a_{n}}\frac{\delta}{\delta R^{a_{1}...a_{n}}}_{s}+\delta s \cdot D_{s}
\end{equation}
where $D_{R}$ is the total derivative with respect to $R$. Such general variations have been used in order to study the stability of the flow. 
Having this formalism we can now analyse different ways in which information regarding the high energy domain can enter the lower energy scales. 
\section{maps of maps}
In general, the renormalisation group is a method that has been used to obtain some information beyond the immediate perturbative regime because the existence of an exactly verified group law that generates the transition between scales allows us to adjust one order of the perturbation expansion with information from higher orders. The fact that the theory can be parametrised according to experimental data at different scales and that the physical observables must produce similar results allows us to express perturbative approximations at one order, using information from higher orders that is included by demanding the exactness of the renormalisation group operation. There are however some implicit simplifying assumptions that are being made. First, the self-similar nature of the renormalisation group means basically that the same type of information is to be found at different scales, once the divergences are eliminated. The prescription of renormalisability basically means that replacing only a finite number of parameters in the theory we can re-define our theory at any scale within the domain of validity of our effective description. Those requirements restrict the type of information that we can transfer across scales and amounts to ignoring various possible effects of high energy physics. It is not only a matter of the physics beyond the cut-off that could influence the low energy physics, but indeed it is a matter of the physics of any given scale below the cut-off being drastically simplified to a self-similar construction. While such an approximation works well in the simplest of the condensed matter problems, if we want to describe the molecular dynamics of more complex systems, leading for example to living systems, such a self-similar approach to function and activity at every scale is not tenable. In terms of the functional description, there are several ways in which we may go beyond this problem. As presented above, the hardest is to assume no self-similarity and to just perform ab-initio computations for higher and higher scales of molecular structures. This approach is computationally hopeless. However, it is possible that there exists a simplifying structure that acts over the types of structures that are being connected between scales. 
\par In the standard approach, the regularisation functions are analysed in the so called optimisation theory. There the problem appears either due to perturbative expansions that have to be finite, or due to truncations of either towers of equations or series that are not constructed in terms of parameters that are guaranteed to be small or controllable. In those cases, a truncation can lead to massive errors and complete cancellation of physically important effects. Therefore, one problem is to see how such a truncation can be protected from artificial non-physical truncation based effects by analysing different types of regulator functions. The usual demand is that the optimised flows should lead to results that are as close as possible to the full theory within each order of the truncation we use. However, aside of the control of the truncation errors, we may as well introduce additional information via the regulator function and how it tends to zero in various cases. Such introduction of ad-hoc effects based on Ansatze of various types is of course possible, but I tend to call this approach "postulating" instead of a more natural "exploratory" approach. As opposed to coming from the outside with our own expectations and impose them to the theory, it would be more interesting to device theoretical instruments capable of probing and scanning for new structure that could exist in various scales and in the connections between scales. The flows across scales are based on the existence of a finite Schwinger functional $W$ and finite correlation functions $\mathcal{O}[\phi]$ for the full theory. The regulator function acts via operators on $W$ and $\mathcal{O}$ and result in flows connecting initial conditions to the full theory. The general assumption is that correlation functions and Schwinger functions for the full theory $\mathcal{O}[\phi,R\rightarrow 0]$ do not depend on the regulator. If correlation functions appear as the result of some truncation $\mathcal{O}[\phi,0]$ they may depend on the chosen flow trajectory given by $R(k)$. Such a dependence should not, again, as assumed up to now, appear for full flows. We can imagine a systematic truncation scheme as follows: at each order of the expansion we add new independent operators in the theory, increasing at each step the number of independent correlation functions. Also, at each step, such correlation functions take various regulator dependent values. We define correlation functions $\mathcal{O}[\Phi]$ as either given by $\tilde{I}[\Phi]$ or constructed from it as $\tilde{I}$ includes all moments of the Schwinger functional. Let us write a correlation function $\tilde{I}[\Phi, R]$ in orders of the truncation
\begin{equation}
\tilde{I}_{k}^{(i)}[\phi, R]=\tilde{I}_{k}^{(i-1)}[\phi, R]+\Delta^{(i)}\tilde{I}_{k}[\Phi, R]
\end{equation}
where the contribution of the $i$-th order is given by $\Delta^{(i)}\tilde{I}$. Now we notice that the R-dependence of $\tilde{I}^{(i)}[\Phi, R]$ is once via the functional $R$-dependence, that identifies a single path in the space of all possible paths in theory space, and by means of $k$ which specifies a certain point on that path, which is defined as $\frac{k}{\Lambda}\in [0,1]$. Therefore variations at a given $k$ are local variations at a given scale, while variations of $R$ are variations changing to different paths in the theory space. 
\par Now, the standard assumption is that the full correlation function in the theory
\begin{equation}
\tilde{I}[\Phi]=\tilde{I}_{0}^{\infty}[\Phi,R]
\end{equation}
is not depending on $R$. If such a dependence were present it would appear as some $R$-dependent renormalisation group reparametrization, which would not appear in an RG invariant object. With this assumption the optimisation of a correlation function $\tilde{I}$ at a given truncation order $i$ implies the following minimisation procedure
\begin{equation}
\min_{R(k)}||\tilde{I}[\Phi]-\tilde{I}_{0}^{(i)}[\phi,R]||=\min_{R(k)}||\sum_{n=i+1}^{\infty}\Delta^{(n)}\tilde{I}_{0}||
\end{equation}
over the space of parametric flows $R(k)$. This assumption is incomplete. If we are to admit the horizontal map of maps relation between the various $R(k)$ trajectories and we allow them to also have a group structure associated to them, the minimisation procedure is strongly restricted. First, if we assume a self-similarity in the space of maps, not every transformation 
\begin{equation}
R(k)\rightarrow R(k)+\delta R
\end{equation}
can be allowed in the minimisation procedure. If the connection between various $R(k)$ is to have a group structure, as implied by a categorification of the renormalisation group, then 
\begin{equation}
R(k_{1})\circ R(k_{2})\rightarrow R(k_{1})\oplus R(k_{2})
\end{equation}
and hence the truncation at the level of $\tilde{I}$ will affect the evolution between various regulator functions. Moreover, while most trans-scale structures disappear from a purely renormalisation group approach to the re-scaling procedure, most of them become encoded in restrictions on what kind of transformations between regulator functions can be chosen. As can be seen, the final theory will not depend on the regulator itself, but it will depend on the fact that certain regulators at each truncation level cannot be chosen. The space of possible "theories" therefore has a structure quite different from the smooth metric space we imagined to be. We can assume that smoothness can be recovered after going to the full theory, but this new full theory will make the restrictions in the minimisation prescriptions of the terms in the truncation for $R(k)$ manifest. Of course, we do not have access to the full theory, or else it wouldn't make sense to start a truncation scheme. Therefore the optimisation is usually performed within each order of the truncation scheme or with respect to advanced truncation steps. I should better be careful in this article to distinguish the full theory and the physical theory. The full theory is the theory in all truncation levels, basically the un-truncated theory, which was assumed in previous studies to be also the physical theory. However, as I said, the truncation itself is not the only source of non-physical effects. The assumption that there is no restriction and no additional structure on the space of transformations of regulator functions also leads to the loss of information. Therefore, the introduction of a group structure over those transformations between regulator flows and the existence of such structure, that ultimately will have an effect on the various outputs of the theory, will modify the theory further, and to that more complete theory I will refer as "physical". Therefore, we may call a regulator independence of the physical theory, but then this physical theory must incorporate the real trans-scale and intra-scale effects that make for example the folding of proteins such a difficult process to simulate. This physical theory must also restrict the space of possible regulator transformations, linking one regulator to another. This ultimately physical theory is indeed independent of the regulator, as long as the regulator is chosen to take into account all physical effects measuring the relations between scales. We can express the relation between the variation of $\tilde{I}_{0}$ with respect to the path $R(k)$ as
\begin{equation}
\delta R^{a_{1}...a_{n}k}\frac{\delta \tilde{I}_{0}[\phi, R]}{\delta R^{a_{1}...a_{n}k}}=\delta (ln \mu) D_{\mu}\tilde{I}_{0}[\Phi, R]
\end{equation}
where the right hand side accounts for the RG variant nature of $\tilde{I}[\Phi]$ at $k=0$ via $\delta \mu(R, \delta R)$, $\mu(R)$. 
Assuming that the truncation affects the correlation functions $\tilde{I}$ such that we can consider
\begin{equation}
||\sum_{n=i+1}^{\infty}\Delta^{(n)}\tilde{I}||=\sum_{n=i+1}^{\infty}||\Delta^{(n)}\tilde{I}||
\end{equation}
the minimisation prescription becomes defined as a constraint on $\tilde{I}^{(i)}$ at each truncation order $i$ and will affect each term $||\Delta^{(n)}\tilde{I}||$ separately. This leads to 
\begin{equation}
\delta R^{a_{1}...a_{n} k}\frac{\delta ||\tilde{I}_{0}[\Phi,R]-\tilde{I}_{0}^{(i)}[\Phi,R]||}{\delta R^{a_{1}...a_{n}k}}=0
\end{equation}
If we see the previous equation as an integrability condition for the flow we can write 
\begin{equation}
\tilde{I}_{0}[\Phi, R]=\tilde{I}_{\Lambda}[\Phi, R]+\int_{\Lambda}^{0}\frac{dk}{k}\partial_{t}\tilde{I}_{k}[\Phi, R]
\end{equation}
and using it in the flow relation
\begin{widetext}
\begin{equation}
D_{R}\tilde{I}[\Phi,R]_{R(\Lambda)}+\int_{\Lambda}^{0}\frac{dk}{k}\partial_{t}[D_{R}\tilde{I}[\Phi, R]]_{R(k)}=\delta(ln \mu)D_{\mu}\tilde{I}_{0}[\Phi, R]
\end{equation}
\end{widetext}
The integrand appears as a total derivative and given that $\delta R_{R=0}=\delta \mu$ we obtain an identity. A variation of the initial regulator $R(\Lambda)$ leads to a variation of $\tilde{I}[\phi, R(\Lambda)]$ and this cannot be kept fixed by adjusting the RG scaling. A different momentum dependence of the regulator $R(\Lambda)$ will lead to different composite operators that will couple to the theory via $\Delta S$. 
\par As we eliminate the regulator in the calculation, for example, we obtain an expansion in the power of the propagators 
\begin{equation}
G=\frac{1}{(\Gamma_{k}^{(2)}[\phi]+R)}
\end{equation}
which carriers residual information regarding $R$ even as the elimination of $R$ is taken. Given several observables $\lambda_{m}$, $m=1,...,m_{max}$, at a degree of truncation $i$ given a truncation scheme, we may search for local extrema of the observables $\lambda_{m}$ in the regulator space. As we do not know the physical realisation of the observables, we can only hope to solve the equation
\begin{equation}
\delta R^{a_{1},...,a_{n}k}\frac{\delta \lambda_{m}}{\delta R^{a_{1}...a_{n}k}}=0
\end{equation}
The general method usually employed is to constrain the set of possible regulators by means of fixing the values of some physically known observables say $\lambda_{1},...,\lambda_{r}$ resolving the optimisation problem for the other observables $\lambda_{r+1},...,\lambda_{m_{max}}$ with a reduced set of regulators. This method is however extremely impractical it destroys the predictive power of the model and usually, additional constraints are required. 
In my approach, the constraints should be universal and allow new relations obeyed by the regulator functions, now seen as maps obeying (semi)group like relations resulting from the categorical interpretation. 
The various possible choices of $R$ result in a modification of the structure of the propagator including terms that appear to have a structure inherited from the higher energy UV domain
\begin{equation}
G=\frac{1}{(\Gamma_{1}\delta^{*}R_{1}+\Gamma_{2}\delta^{*}R_{2}+...)}
\end{equation}
where $\delta^{*}R_{i}$ represents the contribution of the context of the higher energy degrees of freedom, when the regulator is removed. Its value appears by including the requirement of a group law between the maps in a categorical sense, and the proper elimination of the cut-off thereafter. 


\section{Future work}
This article emerged from the desire to identify new structure and correlation originating at or beyond the usual cut-off of our theories, the main motivation being that Nature indicates quite clearly by means of the Higgs boson, the hierarchy problem, the cosmological constant, but also the various types of hierarchies found in molecules with biological impact, that hitherto unknown connections between scales must exist. 
The general observation that Nature has a series of "spurious" or "irrelevant" degrees of freedom is usually accepted by most researchers nowadays. However, this seems to generate certain limitations in our understanding of various phenomena, from quantum chemistry and molecular biology to molecular evolution theory. 
In fact, ref. [20] introduces the idea of equivalence classes leading to the conclusion that the "spurious" information (also called inessential) is to be disregarded in favour of the so called "essential" physical observables. My view, underlined also in my early articles [21], [22], is that this type of "redundant" or "spurious" information is in fact not that spurious, and that it can in principle be used to determine additional structure and also to simplify various problems, however, not by ignoring it but by making use of it and trying to find new, hitherto unnoticed correlations within it. 
In this article I introduced the concept of cooperative behaviour instead of the well known collective behaviour. 
This concept appeared in biology and I found it inspiring while thinking about the hierarchy problem in physics. As opposed to the behaviour of electrons in a metal or of other collective phenomena, in which large numbers of particles have qualitatively similar behaviour, biology behaves by quite different rules. The constituent element of a biological structure is a cell, and a cell has its own internal functions that allow it to live, and in extreme situations, even to live independent from the larger organism. When however, such a cell enters a complex environment like that of the body of a multi-cellular organism, it adapts its inner workings to the requirements of the organism, while not completely abandoning its original capabilities. The evolution of the cell and its reproduction are controlled by a series of higher scale mechanisms to which the cell itself adapts its inner structure. Clearly we do not have a collective behaviour anymore, like the one we would find in condensed matter systems, but instead, we have what I call a cooperative behaviour in which different scales have various means of communicating with each other and of "translating" information at one scale such that it has a more or less precise effect at a different scale. Biologists have observed such effects when trying to provide mechanisms for example for embryogenesis, or for the foetal development of the nervous system. The process is far from being completely understood, and much of the effects biologists observe at those stages are not explained by identifiable mechanisms. However, it seems clear that a high level of inter-scale communication and influences must play a role in those processes. Clearly, biologists will most likely pinpoint a series of complex biochemical processes going on at these levels, but I suspect that the underlying mathematical structure is of the form of a categorical renormalisation group. This structure provides us with a subtle form of correlation and communication between scales that cannot be observed by looking only through the lense of the classical renormalisation group. 

 I also noticed that the concept of "emergence" is often misused (from the perspective of a physicist) giving it properties that, as we know from physics, it does not possess. It is hard to claim, for example, that "consciousness emerges from neural cells" in the same way plasmons "emerge" from the collective fluctuations of conduction electrons. In fact, in biology, each scale has a series of feedback relations with other scales, allowing for adaptation at each scale. The formation of multi-cellular complex organisms like animals is a clear example of that. However, why and how such inter-scale mechanisms formed is not clear and it is an active subject of research in molecular evolution theory [23]. Even if the molecular mechanisms underlying such informational feedback between the scales would be understood, and biologists are far from that, there is still the question as of why evolutionary processes led to such mechanisms? Is there an underlying mathematical principle that allows for that? If we are to think only in terms of the renormalisation group as understood now, there is little chance to understand such a mechanism. A purely collective behaviour as taught by the renormalisation group approach alone, would not allow for such complex inter-scale relations. The existence of such an inter-scale feedback in which global information about large scale structures is somehow made available (or translated) to the lower scale cells or individuals and the fact that in this process, neither of the constituents completely loses its individual ability to function independently [24] made me look for a more fundamental mechanism for such a connection between scales. As made clear by [24], a cell taken out of an organism can reverse to its "wild" variety, regaining its ability to function and reproduce independently. Therefore, the instructions as of how to behave as an independent cell are not eliminated, but are adapted to the requirements of the organism when the cell lives within the organism. This situation in which the small scale degrees of freedom are not simply averaged over, but in which they keep their independent behaviour which is incorporated in a very well controlled way in the higher functions of the organism is what I call "cooperative behaviour". A strictly emergent collective behaviour, a la conventional renormalisation group, of, say, liver cells, would lead to cancer. 
I suspect that the cooperative mechanism that lies at the foundation of the evolution processes in biology is based on this categorical extension of the renormalisation group, as presented here. Further research is of course required to confirm or infirm that. As I am not well aware of various progresses in molecular biology or molecular evolution theory, I am currently working on another problem where the mechanisms of the categorical renormalisation group may play a fundamental role. 


\par
One problem of quantum chemistry is the difficulty of treating strongly correlated electronic systems. The correlation between electrons in materials (be it condensed matter or molecular systems) is fundamental in describing its properties. While in condensed matter, the system is well described by a renormalisation group approach, the application of this principle to quantum chemistry only came later, and the introduction of the so called density matrix renormalisation group was a key step in the process. In quantum chemistry, usually, the description of correlation is performed by the so called multi-reference configuration interaction method (MRCI). This method involves the introduction of a series of different slater-determinant type configuration functions from which various swaps are being performed in order to create a model that better approximates the strong correlations in quantum chemical systems. However, the main problem is that this approach grows exponentially in complexity with the number of orbitals and electrons involved, making the MRCI approach unsuitable for larger molecules and in particular for situations in which one wishes to describe, for example, the way a molecule fragment at a given scale is related in function or geometry with another molecule at a larger scale. There are various situations in which strong correlations appear in quantum chemistry situations. The so called static correlation is a correlation that appears due to the (cvasi)degeneracy of molecular orbitals in certain contexts (for example in the context of near dissociation compounds). The other correlation appears when the gap between the highest occupied molecular orbital and the lowest unoccupied molecular orbital is large enough for the wavefunction to be dominated by the Hartree Fock reference determinant. In this case, we have the so called dynamical correlation, which becomes dominant in the small distance region where it can not easily be described by current methods. 
The standard implementation of the density matrix renormalisation group has been able to deal relatively fine with complicated strong statical correlation problems for molecules that are more or less linear. The reason why non-linear molecules are not easily treated in this sense is mainly that the DMRG method automatically enforces locality by means of the correlations it induces progressively in the description, but only in one dimension. The method functions by introducing auxiliary indices on the coefficients of the wavefunctions, such that they separate into a set of matrices multiplied with each other according to sum rules following those auxiliary indices. This multiplication together with the DMRG blocking and sweeping ansatze not only accurately describes the correlation, but also reduces the correlation complexity when various orbitals are linked together, in a way that insures locality. The method is based on a linear orbital setup. This is because the matrices depending on auxiliary indices do not commute and this leads to two problems: not only do we have to carefully choose the individual orbitals involved in the chain, but also their ordering, demanding that the respective groupings of matrices can properly take into account physical correlation. While in a linear molecular structure, the ordering that enforces locality is natural and a RG method can easily be implemented, when a non-linear structure is introduced, DMRG fails because the introduction of locality in the other directions amounts to exponential complexity growth in the types of correlations to be considered. In fact, the DMRG performs a simplification of the types of correlations that encode locality in only one dimension, leaving the other dimensions unresolved. The dynamical aspect of the correlation in DMRG is related to the fact that a proper regularisation was always assumed but never actually performed. The first situation in which some kind of regularisation was introduced was in [25] where the small scale behaviour was described in terms of density functional theory. However, this approach shifts the treatment of small scale, large correlations in the realm of semi-empirical methods. Therefore, again, we know there must be some structure at the small distance scale, but that structure is never captured by the DMRG method, leading to problems with the treatment of molecular systems that strongly depend on dynamical correlations, and, due to a bad small to large scale connection, to the exponential growth in complexity when treating non-linear molecules. 
I expect that the Density Matrix Renormalisation Category will solve these problems or at least alleviate them substantially. While the details are to be discussed in an article that is now under preparation, the main idea is that we assign an additional auxiliary index structure but as opposed to the usual attempts starting with [26], the mappings between regulator functions will be endowed with a group structure describing their transformations. Not only do we have to add more indices to the coefficients, leading to higher tensor forms (as has been done in previous attempts towards multi-dimensional DMRG), but the same indices will be coupled to new matrices that will play the role of variational transformations between different low energy regulator functions. The variational solution of this problem detects structures that were not visible before because they represent correlations in the small scale degrees of freedom that impact the large scale behaviour, hence a "cooperative" behaviour, and that allow for a more flexible variational approach. All other aspects of the DMRG method remain unchanged, with the observation that in the case of DMRC (density matrix renormalisation category) the additional matrices linking different regularisation functions in the "UV" are required to obey a group law. 
\section{conclusion}
In this article I presented an innovative way of looking at the renormalisation group, from the perspective of category theory, modifying its usual interpretation in a non-trivial way. As opposed to considering the regulator functions as purely arbitrary, they become part of a group of maps, in which information is encoded about relations between scales as well as additional information originating in the high energy degrees of freedom that are being integrated over. While the usual method of regularisation eliminates the effect of such degrees of freedom uniformly, here, a group law and hence a hierarchy structure is included due to the categorical approach. Because of this "horizontal" view, we expect such a method to be applicable both to various problems related to biology, like protein folding, etc. as well as to the hierarchy problem.

\end{document}